%% file: main.tex
\documentclass[conference]{IEEEtran}

\usepackage{booktabs}
\usepackage{tabularx}
\usepackage{amsmath}
\usepackage{amssymb}
\usepackage{algorithm}
\usepackage{algpseudocode}
\usepackage{hyperref}
\usepackage{graphicx}
\usepackage[caption=false,font=footnotesize]{subfig}
\usepackage{multirow}

\algrenewcommand\algorithmicrequire{\textbf{Input:}}
\algrenewcommand\algorithmicensure{\textbf{Output:}}

\begin{document}

\title{JANUS: Denial-of-Service Attack Against Beam Hopping in LEO Satellite Networks}

\author{
    \IEEEauthorblockN{
        Yuval Aviv\textsuperscript{*},
        Roee Idan\textsuperscript{*},
        Roy Peled,
        Asaf Shabtai,
        and Yuval Elovici
    }
    \IEEEauthorblockA{
        Stein Faculty of Computer and Information Science\\
        Ben-Gurion University of the Negev, Israel\\
        \textsuperscript{*}Equal contribution
    }
}

\maketitle

\begin{abstract}

Low Earth orbit (LEO) satellite networks are increasingly used to provide global connectivity.
However, each satellite has limited resources that need to be allocated according to demand, which varies geographically and over time.
Beam hopping addresses this challenge by dividing a satellite’s service area into geographic cells.
Rather than illuminating every cell simultaneously, it dynamically assigns available beams to a selected subset based on demand.
This reliance on observed traffic demand as an input to beam-selection decisions creates a new attack surface whose security implications have received little attention.
In this paper, we present JANUS, a novel targeted denial-of-service attack against beam-hopping systems in LEO networks.
We show that a small botnet of compromised terminals can inject legitimate user traffic into carefully selected non-victim cells to manipulate the beam-hopping scheduler's view of demand.
This manipulation alters beam-allocation decisions and redirects service away from the targeted victim area.
We evaluate JANUS across different system configurations, schedulers, attack horizons, and attacker-knowledge settings to characterize the attack’s effectiveness, required resources, and resulting service disruption over time.
Against a rank-based KMAX scheduler, JANUS achieves complete service denial for up to approximately 95\% of evaluated victims.
Against DRL, JANUS can exclude the victim from approximately 92\% of scheduling decisions.
Finally, we evaluate mitigation strategies that reduce the attack effectiveness.

\end{abstract}

\IEEEpeerreviewmaketitle

\input{sections/Introduction}
\input{sections/Background}
\input{sections/ThreatModel}
\input{sections/JANUS_Attack_Framework}

\input{sections/Results}
\input{sections/Mitigation}
\input{sections/Discussion}
\input{sections/Conclusion}
\input{sections/Ethical}

\bibliographystyle{IEEEtran}
\bibliography{references}

\input{sections/Appendix}

\end{document}

%% file: sections/Introduction.tex
\section{Introduction}
\label{sec:introduction}

In recent years, Low Earth orbit (LEO) satellite networks have progressed from emerging systems toward deployed broadband infrastructure, driven by large commercial constellations such as SpaceX's Starlink, Eutelsat OneWeb, and Amazon Leo~\cite{starlink,oneweb,amazonleo}.
These networks are also expected to play an integral role in future 6G systems, extending broadband connectivity to remote, underserved, and highly mobile users beyond the reach of terrestrial infrastructure~\cite{ntontin2025vision}.
Unlike terrestrial networks, the topology and coverage of LEO satellite networks change continuously as satellites move relative to Earth.
As satellites move, the set of ground areas each satellite serves changes, while user demand remains uneven across geographic areas and varies over time~\cite{lei2024spatial}.
Allocating resources across changing coverage and uneven demand is made more difficult by the limited power, spectrum, and beamforming resources available to each satellite, which restricts the number of ground areas that can be served simultaneously~\cite{lin2022multi}.
Efficient LEO network operation must therefore match limited and time-varying satellite resources to changing demand.

Beam hopping (BH) provides a mechanism for allocating limited satellite resources across a large coverage area~\cite{guo2022efficient,zhang2023system}. 
Rather than illuminating every cell simultaneously, BH allows a satellite to illuminate a subset of cells in each decision window and reallocate its beams as demand changes~\cite{guo2022efficient,zhang2023system}.
This allows the system to concentrate its available power, spectrum, and bandwidth on cells with higher service demand, rather than allocating beam time to cells with relatively low demand~\cite{guo2022efficient,lin2022dynamic}.

Dynamic BH schedulers use traffic and network state, including cell demand, queue occupancy, and channel conditions, to select beam patterns and allocate bandwidth so that available resources better match demand~\cite{lin2022dynamic,chen2025distributed}. 
Recent studies have applied deep reinforcement learning (DRL) to learn these allocation policies over successive decision windows, allowing the scheduler to adapt beam patterns and associated power and bandwidth allocations as traffic and network conditions evolve~\cite{lin2022dynamic,xu2023novel,chen2025distributed}. 

Demand-responsive BH introduces a new attack surface through its reliance on observed traffic demand.
In dynamic BH, this demand directly affects which cells are selected for illumination in subsequent decision windows~\cite{lin2022dynamic,chen2025distributed}.
An adversary that manipulates the scheduler's view of demand can therefore influence its beam-allocation decisions.
To manipulate the scheduler, the adversary can generate legitimate user traffic in carefully selected non-victim cells, making those cells appear more urgent than the target victim cell and redirecting service away from it.

We present JANUS, a novel targeted denial-of-service (DoS) attack against the BH scheduler of a satellite in a LEO network.
JANUS targets a geographic area within a satellite's footprint by manipulating the scheduler's view of demand, causing the cell serving the targeted victim area to be excluded from beam selection while competing cells are selected instead. 
The attacker uses a distributed set of compromised user terminals to generate coordinated traffic in these non-victim cells. 
This traffic inflates their apparent demand and causes the scheduler to prioritize them over the victim cell. 
As a result, users in the targeted area experience reduced service quality or are denied service altogether during the attack.

Prior research on LEO DoS attacks has shown that the predictable evolution of satellite topology and routing, along with uneven traffic and limited link capacity, gives rise to time-varying bottlenecks across ground-satellite and inter-satellite links~\cite{giuliari2021icarus,deng2025time}.
These attacks exploit such bottlenecks by steering botnet traffic through selected links until they become congested.
JANUS targets a different layer of the system.
Rather than congesting links or disrupting routes, it manipulates the demand observed by the BH scheduler, causing beam resources to be redirected away from a target geographic area.
Unlike prior attacks that target network links and routing, JANUS targets the scheduling layer, exposing a new attack surface in LEO satellite networks.

We evaluate JANUS against both rank-based and learning-based BH schedulers to examine its effectiveness against explicit demand-ranking rules and adaptive policies learned over successive decision windows.
We evaluate both single-window and multi-window attacks, where each window represents the time interval between consecutive beam-allocation decisions.
During multi-window attacks, each beam-allocation decision changes the queue state observed in subsequent windows, so the attacker must account for how earlier attack decisions affect later scheduling decisions.
We evaluate the attack's effectiveness and resulting service degradation across constellation profiles, BH configurations, schedulers, attack horizons, and attacker-knowledge settings, and characterize the traffic and botnet resources required to carry out the attack.

We implement JANUS by extending the ICARUS simulator~\cite{giuliari2021icarus} with dynamic BH scheduling and adversarial traffic injection.
JANUS excludes the victim in $98.7\%$ of single-window attacks against KMAX and achieves full-horizon exclusion for $94.8\%$ of victims over multi-window attacks.
Against DRL, JANUS achieves approximately $67$--$92\%$ single-window attack success rate and reduces the amount of data received by the victim by at least $77$--$81\%$ over multi-window attacks.
These results show that JANUS can sustain substantial service disruption across a range of system and attacker conditions.

We also evaluate several defenses that aim to reduce the effectiveness of JANUS while retaining the scheduler's ability to respond to legitimate demand changes.
These mechanisms include randomized and starvation-aware beam reservation, limits on consecutive service, and smoothing of the demand observed by the scheduler.
Our results show that scheduler-side defenses can substantially reduce JANUS effectiveness, with their impact varying across scheduler designs.

Our contributions are as follows:
\begin{itemize}
    \item We identify demand manipulation in demand-responsive BH as a security-critical attack surface that underlies a new class of targeted scheduling-layer DoS attacks in LEO satellite networks.
    
    \item We present JANUS, a novel targeted DoS attack against the BH scheduler that redirects service away from a targeted victim area using a small botnet of compromised terminals, generating legitimate user traffic individually indistinguishable from legitimate demand.

    \item We extend ICARUS simulator~\cite{giuliari2021icarus} to support dynamic BH scheduling with rank-based and learning-based schedulers, providing a reusable platform for BH security research.
    
    \item We propose and evaluate scheduler-side mitigations for demand manipulation, examining their effectiveness in reducing attack impact.
\end{itemize}

To the best of our knowledge, JANUS is the first targeted scheduling-layer DoS attack against BH in LEO satellite networks that manipulates observed demand to divert beam resources away from a target geographic area.
In addition to demonstrating the feasibility of the attack itself, our work identifies demand-driven beam allocation as a security-critical attack surface and provides a basis for the design and evaluation of more robust BH schedulers.


%% file: sections/Background.tex
\section{Background and Related Work}
\label{sec:background}

LEO satellite networks must serve geographically uneven and varying traffic demand using limited power, spectrum, and bandwidth.
BH addresses this challenge by dynamically concentrating resources on selected service cells.
This section describes the LEO network characteristics and BH scheduling mechanisms relevant to JANUS, and reviews related work on LEO-network DoS attacks and adversarial manipulation of adaptive decision-making systems.

\subsection{LEO Satellites and LEO Satellite Networks}

LEO satellites operate at altitudes below 2,000 km and complete an orbit around the Earth in roughly 90 minutes~\cite{darwish2022leo}.
In contrast, geostationary Earth orbit (GEO) satellites operate much farther from Earth, at an altitude of 35,786 km, and appear fixed over a location on the ground~\cite{ntontin2025vision}.
The lower orbital altitude of LEO satellites reduces propagation latency compared to GEO systems, but LEO satellites move continuously relative to users on the ground, cover only a limited geographic area at any given time, and remain visible from a given location for only a limited period~\cite{darwish2022location,zhang2022enabling}.
Spacecraft and launch constraints impose strict mass and volume limits, while onboard power and thermal budgets further restrict the resources available to each satellite~\cite{sauder2024framework,diaconu2024phase}.
These constraints limit transmit power and onboard processing capacity, while finite bandwidth and spectrum allocations further constrain aggregate communication capacity~\cite{bhandari2024user,heydarishahreza2024spectrum}.

To address the limited coverage of individual satellites, LEO satellite networks use constellations of hundreds to thousands of satellites to provide continuous wide-area connectivity~\cite{darwish2022leo,lai2024your}.
The satellites communicate with user terminals and gateways through ground-to-satellite links (GSLs) and with other satellites through inter-satellite links (ISLs)~\cite{ntontin2025vision}.
As the satellites move, their coverage and the available GSLs change continuously, producing a dynamic network topology~\cite{zhang2022enabling,deng2025time}.

User terminals connect to a serving satellite through a GSL and are handed over to another satellite as coverage changes~\cite{darwish2022location}.
Depending on the constellation architecture and deployment stage, traffic can reach a gateway directly from the serving satellite or traverse multiple ISLs first~\cite{zhang2022enabling}.
Gateways connect the satellite constellation to terrestrial points of presence and the wider internet~\cite{darwish2022leo}.
The scale of these networks, their continuously changing coverage, and their limited satellite resources make efficient resource allocation essential~\cite{bhandari2024user}.

\subsection{Beam Hopping}

BH is a satellite resource-allocation mechanism in which a satellite illuminates only a subset of the cells in its service area during each decision window~\cite{zhao2022beam}.
This technology was initially developed for GEO systems, whose large footprints often include low-demand regions such as oceans and deserts, making continuous uniform illumination inefficient~\cite{anzalchi2009beam}.
In recent years, BH has been adapted to LEO satellites to obtain similar benefits by directing limited resources toward cells with greater demand~\cite{guo2022efficient}.
This application of BH to LEO has attracted increasing research interest and has also been demonstrated in orbit, for example through Eutelsat OneWeb's JoeySat~\cite{shuang2024joint,esa2025joeysat}.
Under BH, the satellite concentrates its limited power and bandwidth on selected cells instead of continuously serving the entire footprint~\cite{li2021overview}.

\begin{figure}[htb]
    \centering
    \includegraphics[width=\columnwidth]{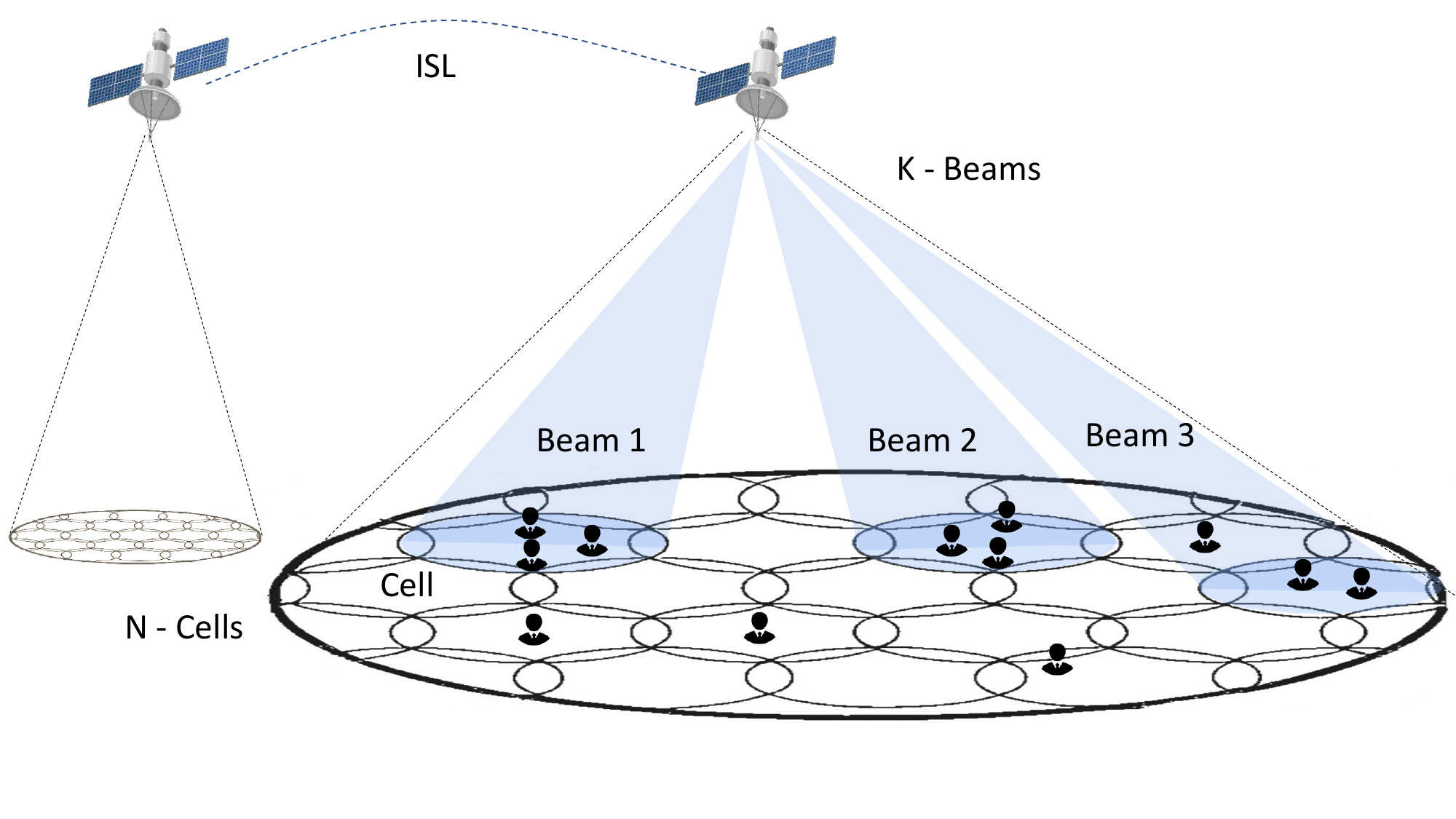}
    \caption{BH operation under benign conditions. A satellite footprint is partitioned into $N$ candidate cells, while only $K<N$ cells can be illuminated simultaneously in each decision window. The scheduler selects which cells receive the available beams based on the current network state, including their observed traffic demand.}
    \label{fig:bh-overview}
\end{figure}

As illustrated in Figure~\ref{fig:bh-overview}, for BH scheduling, the satellite coverage footprint is represented as a set of geographic service cells, each of which can be selected for illumination~\cite{zhao2022beam}.
In each decision window, the BH scheduler decides which of these cells to illuminate.
If a satellite has $N$ candidate cells and can illuminate at most $K$ cells simultaneously, where $K<N$, then at least $N-K$ cells are not illuminated during that decision window.
BH scheduling is therefore a resource-prioritization problem.
Existing BH designs commonly use traffic-derived state to guide this prioritization, including queue length, recent arrivals, estimated or predicted demand, channel-aware service estimates, and learned traffic features~\cite{zhao2022beam,li2021overview}.
Under benign conditions, demand-responsive scheduling directs limited satellite resources toward cells with higher observed demand~\cite{lei2024spatial,lin2022dynamic,chen2025distributed}.

\subsubsection{Dynamic Beam Allocation}

Dynamic beam allocation allows the scheduler to adjust beam-illumination patterns and the bandwidth and power assigned to the selected beams in response to changes in traffic demand and channel conditions.
Classical approaches formulate this as an optimization problem involving beam-pattern selection, user scheduling, bandwidth allocation, and power allocation~\cite{zhao2022beam,li2021overview}.
Lei et al.~\cite{lei2024spatial} jointly optimized beam-pattern selection, user association, user scheduling, and power allocation under spatially and temporally uneven traffic, showing improved matching between offered capacity and user demand, while Lin et al.~\cite{lin2022multi} coordinated BH patterns across multiple satellites to balance traffic while reducing intra-satellite and inter-satellite interference, improving load balance and increasing served traffic.
Yuan et al.~\cite{yuan2026radio} further considered traffic and interference-aware BH scheduling, improving throughput and user-demand satisfaction over greedy allocation.

The large and sequential decision spaces associated with dynamic beam allocation have motivated the use of learning-based approaches.
In another study, Lin et al.~\cite{lin2022dynamic} used cooperative multi-agent deep reinforcement learning (MADRL) for joint dynamic beam-pattern and bandwidth allocation, showing that the learned policy could support real-time scheduling under non-uniform and time-varying traffic demand.
Chen et al.~\cite{chen2025distributed} formulated distributed BH scheduling as a multi-agent learning problem in which satellites use local channel and queue information, achieving higher throughput and lower transmission delay than centralized baselines.
Using proximal policy optimization with a hybrid action space, Xie et al.~\cite{xie2025multi} jointly select discrete illumination patterns and continuous power allocations over time, maintaining high throughput while reducing delay.
In related work performed by Liang et al.~\cite{liang2024multiagent} multi-agent DRL was applied to two-timescale bandwidth allocation in multibeam satellite networks, reducing communication delay and improving fairness compared with a single-agent DRL approach.
More recently, Zhang et al.~\cite{zhang2025efficient} used a multi-agent actor--critic approach for joint BH-pattern and power allocation in LEO networks, dynamically allocating power according to traffic demand while improving throughput and latency.

Classical and learning-based schedulers differ in terms of how they produce allocation decisions.
While classical schedulers use explicit optimization procedures or priority rules, and learning-based schedulers map multidimensional network state to illumination and resource-allocation actions~\cite{li2021overview,lin2022dynamic}, all of the approaches mentioned above rely on traffic-derived state.
This shared reliance on traffic-derived state improves resource allocation under benign conditions but also exposes the scheduling process to adversarial demand manipulation.

\subsection{Denial-of-Service Attacks Against LEO Networks}

The radio, ground, user, and network segments of satellite communication systems face availability threats, including jamming, terminal compromise, malware-enabled disruption, and traffic flooding~\cite{salim2024cybersecurity}.
In LEO networks, global accessibility, predictable satellite motion and routing, and constrained link capacity create additional opportunities for coordinated network-layer DoS attacks~\cite{giuliari2021icarus,deng2025time}.

Terrestrial link-flooding attacks such as Coremelt~\cite{studer2009coremelt} and Crossfire~\cite{kang2013crossfire} showed that legitimate user traffic generated through compromised terminals can be directed through selected links to congest network infrastructure without directly flooding the victim endpoint.
ICARUS~\cite{giuliari2021icarus} extends this attack model to LEO satellite networks by using compromised satellite-enabled hosts to generate coordinated traffic whose routes converge on selected GSLs or ISLs.
The attack exploits the global accessibility, predictable topology, limited link capacity, and constrained path diversity of LEO networks.

Subsequent work examined other ways in which LEO topology dynamics can support DoS attacks.
Lu et al.~\cite{lu2024dosat} proposed DoSat, which exploits routing changes during topology transitions to concentrate attack traffic on selected links.
STARMAZE, introduced by Wang et al.~\cite{wang2024starmaze}, instead targets selected ISLs to produce persistent routing detours and service degradation.
More recently, Deng et al.~\cite{deng2025time} presented SKYFALL, which identifies time-varying bottleneck GSLs and uses coordinated traffic to congest them and reduce throughput.
HYDRA~\cite{idan2026hydra} further examines LEO link-flooding attacks by quantifying the botnet resources required to effectively target different network links.

Unlike these forwarding-layer attacks, JANUS targets the scheduler by manipulating the traffic-derived state used to allocate beam resources.
The demand-responsive behavior of BH makes this attack different from ordinary traffic flooding and exposes a previously unexamined attack surface.
JANUS therefore connects prior work on coordinated traffic-based DoS with adversarial manipulation of decision-making systems.

\subsection{Adversarial Manipulation of Scheduling State}

Prior work has shown that DRL policies can be influenced through adversarial observations and policies that alter the state or interactions used to select actions~\cite{huang2017adversarial,lin2017tactics,gleave2019adversarial,wang2023adversarial}.
Broader studies have examined these vulnerabilities and their potential defenses across learning-based control systems~\cite{ilahi2021challenges,chen2019adversarial,ren2020adversarial}.
Collectively, these studies demonstrated that an attacker can influence control decisions by shaping the information available to a policy without directly modifying the policy itself.

This makes JANUS an instance of scheduling-layer environment manipulation.
The attacker does not directly attack the victim path, the satellite hardware, or the scheduler implementation.
It changes the environment observed by the scheduler so that normal demand-responsive behavior produces an adversarial allocation.
This view applies to both rank-based schedulers and learning-based schedulers, because both rely on traffic-derived state to decide which cells receive service.

\subsection{Research Gap and Distinction of JANUS}

Prior BH research has primarily evaluated resource allocation under benign traffic conditions, focusing on throughput, delay, load balancing, beam selection, power allocation, bandwidth allocation, and resource efficiency~\cite{zhao2022beam,lin2022dynamic,chen2025distributed,liang2024multiagent,xie2025multi}, while prior studies on LEO DoS attacks examined attacks against the forwarding layer, including routes, ground-to-satellite links, ISLs, and time-varying network bottlenecks~\cite{giuliari2021icarus,lu2024dosat,wang2024starmaze,deng2025time}.
Recently published work (July 2026) by Zheng et al.~\cite{zheng2026beam} evaluates the robustness of a DRL-based LEO BH and resource-allocation framework under adversarial perturbations to scheduler-visible link-gain information.
This work targets the state of a learned scheduler, whereas JANUS manipulates the traffic demand observed by the scheduler through legitimate traffic generated by compromised terminals and applies to both learning-based and rank-based BH schedulers.
Despite this extensive research, the security implications of manipulating the traffic demand used by a BH scheduler remain largely unexplored.

To the best of our knowledge, JANUS is the first targeted scheduling-layer DoS attack against LEO BH that manipulates scheduler-visible demand using legitimate user traffic generated through compromised terminals.
JANUS moves the DoS target from the forwarding layer to the scheduling layer.
It does not require compromising satellites, gateways, routing protocols, or the scheduler implementation.
Instead, coordinated traffic sent to selected non-victim cells changes the demand observed by the BH scheduler, causing it to redirect illumination away from a targeted victim cell.
JANUS therefore exposes a previously unexamined attack surface in demand-responsive satellite resource allocation and extends existing work on LEO DoS beyond attacks against links and routes.

%% file: sections/ThreatModel.tex
\section{JANUS Threat Model}
\label{sec:threat-model}

This section defines the threat model considered by JANUS.
We first describe the adversary's objective and the conditions under which the attack is considered successful, and then specify the adversary's capabilities, knowledge, and operational constraints.

\subsection{The adversary's objective}
\label{subsec:adversary-objective}

\begin{figure}[htb]
    \centering
    \includegraphics[width=\columnwidth]{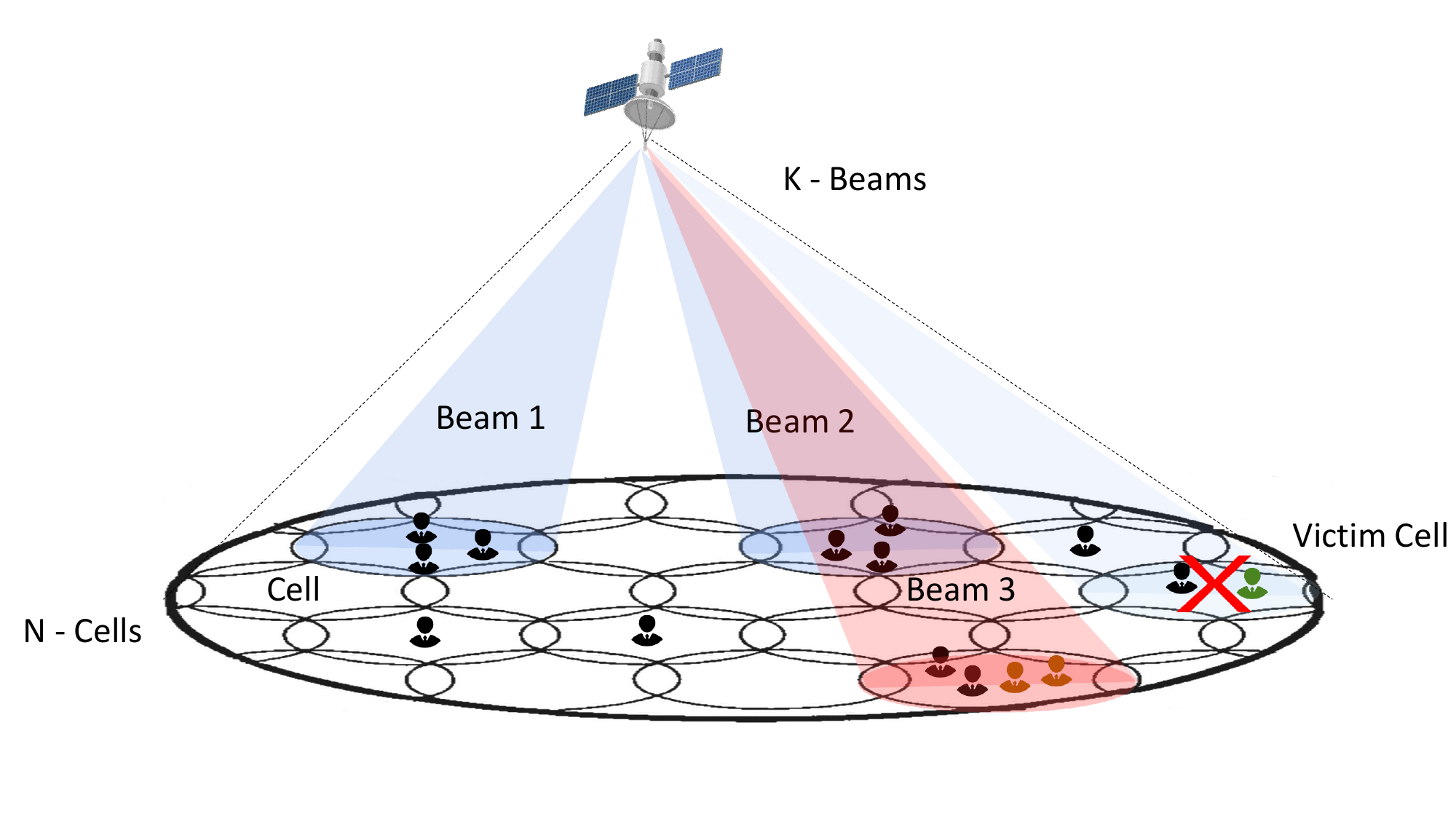}
    \caption{Illustrative JANUS threat model applied to the BH operation shown in Figure~\ref{fig:bh-overview}. The green user is the victim, while the red users are destinations in competing non-victim cells. Compromised terminals generate traffic toward these cells, increasing their downlink demand and causing the scheduler to redirect beam resources away from the victim. The red beam indicates the redirected allocation.}
    \label{fig:janus-threat-model}
\end{figure}

The goal of the JANUS adversary is to create targeted service degradation in a specific geographic area served by a BH LEO satellite network.
The attack manipulates beam-scheduling decisions so that the cell serving the victim area receives insufficient service, or is excluded from beam selection, over one or more consecutive decision windows.
Rather than sending traffic directly to the victim cell, the attacker increases apparent demand in non-victim cells, causing the scheduler to allocate beam resources away from the target, as illustrated in Figure~\ref{fig:janus-threat-model}.

The attack objective has three dimensions.
\begin{itemize}
    \item \textit{Victim cell degradation.}
    The attacker aims to reduce or block service to the BH cell that serves the targeted geographic area.

    \item \textit{Targeted geographic impact.}
    The attacker focuses disruption on users located in a specific victim area, rather than causing broad network-wide degradation.

    \item \textit{Targeted attack horizon.}
    The attack horizon may consist of a single decision window or multiple consecutive decision windows, allowing the attacker to cause degradation or outage during a chosen period.
\end{itemize}

We consider JANUS successful in a decision window when the victim cell is excluded from beam selection.
We evaluate the resulting service impact through reduced throughput, increased queueing delay, and temporary service loss for users in the targeted area.

\subsection{Attacker Capabilities and Constraints}
\label{subsec:attacker-capabilities}

JANUS relies on information that is publicly available or can be estimated from public sources, together with a limited set of compromised traffic sources. 

\begin{itemize}
    \item \textit{Satellite network topology and routing.}
    Satellite orbital information is publicly available through Two-Line Element (TLE) data, and standard orbit propagation models can be used to estimate satellite positions over time~\cite{celestrak_tle,kelso2007validation}.
    From this information, the adversary can construct an approximate time-varying view of the satellite network, including the satellites that are likely to be relevant to the victim area during the attack horizon~\cite{bhattacherjee2019network}.
    For routing, the adversary assumes that traffic follows a deterministic low-latency policy, such as shortest-path or nearest-serving-satellite routing, which is commonly used as a baseline model for LEO network analysis~\cite{bhattacherjee2019network,giuliari2021icarus}.

    \item \textit{Antenna pattern and coverage.}
    The adversary can use publicly available technical reports, regulatory filings, and launch information to infer antenna characteristics, such as beam count, frequency bands, coverage footprint, and service constraints ~\cite{fcc_starlink1}.
    From these characteristics, the adversary can estimate the BH cell layout used within a satellite footprint~\cite{fcc_starlink2}.
    Given this estimated cell layout together with the satellite positions derived from TLE data, the adversary can approximate which ground cells are covered by the relevant satellite during the attack horizon and which non-victim cells may compete with the victim cell for beam service.
    
    \item \textit{Botnet resources.}
    The adversary controls a limited set of compromised user terminals, or user devices connected through such terminals, that can transmit traffic in the satellite network.
    Each compromised terminal acts as an attack traffic source and is constrained by a bounded transmission rate.
    The attack is therefore limited by the number of available compromised terminals, their geographic locations, and their ability to generate traffic over the attack horizon.
    As in prior LEO botnet threat models, the adversary can coordinate these terminals through a command-and-control channel and issue attack commands in advance~\cite{giuliari2021icarus}.

    \item \textit{Traffic-demand distribution.}
    The adversary has an estimate of how traffic demand is distributed across the BH cells.
    Such an estimate can be obtained by passively monitoring satellite transmissions, whose downlink activity varies with user traffic demand~\cite{gomez2024starlink}.
    Beam activity and switching behavior can also be inferred from received satellite signals~\cite{neinavaie2022unveiling}, while passive timing measurements can reveal characteristics of individual beam transmissions and their service intervals~\cite{qin2025timing}.
    By accumulating these observations over time, the adversary can estimate recurring demand patterns across geographic service cells.
\end{itemize}

The adversary is subject to several constraints.
First, the adversary does not compromise the satellite, the BH scheduler, the gateway, ground stations, or the operator's control infrastructure.
Therefore, the adversary does not have access to privileged scheduler state, control-plane commands, routing-table updates, or satellite operational systems.

Second, the adversary cannot directly control beam allocation.
The BH decisions remain under the control of the legitimate scheduler, and the adversary cannot choose which cells are illuminated, modify the beam pattern, or change the physical-layer parameters of the antenna.
The adversary can only influence the scheduler indirectly by changing the traffic demand observed by the scheduler.

Finally, the adversary does not jam the wireless channel, spoof satellite control messages, tamper with legitimate user traffic, rely on malformed packets, or gain direct access to the scheduler.
Instead, JANUS operates through nominal user-generated flows from compromised terminals or user devices, using ordinary demand as the signal that influences beam-scheduling decisions.

%% file: sections/JANUS_Attack_Framework.tex
\section{The JANUS Attack Framework}
\label{sec:janus-framework}

This section presents JANUS, a framework for targeted scheduling-layer DoS against demand-responsive BH.
For each decision window, JANUS identifies the satellite and BH cell serving the victim region, uses its estimate of the scheduler inputs to determine how attacker-generated traffic in selected non-victim cells may affect the illumination decision, and realizes the resulting traffic allocation through nominal traffic generated by compromised user terminals.

JANUS operates at two levels.
At the cell level, it derives an adversarial demand allocation that changes the scheduler's illumination decision.
For rank-based schedulers, JANUS computes the additional demand required under the scheduler's priority rule.
For learning-based schedulers, it searches for a successful allocation within a fixed traffic budget using an evolutionary procedure guided by a surrogate model that approximates the target scheduler.
At the network level, JANUS maps the selected cell-level allocation to feasible source--destination flows subject to compromised-terminal locations and per-terminal uplink limits.
JANUS influences the scheduler through the traffic-derived state it processes, without directly interacting with the scheduler itself.
The scheduler continues to operate normally, but the resulting state may lead it to allocate its limited illumination resources to attacker-influenced cells rather than to the victim cell.
Table~\ref{tab:janus-notation} summarizes the notation used throughout this section.

\begin{table}[t]
    \centering
    \caption{Notation used in the JANUS attack framework.}
    \label{tab:janus-notation}
    \small
    \begin{tabularx}{\columnwidth}{@{}lX@{}}
        \toprule
        Symbol & Meaning \\
        \midrule

        $N$
            & Number of BH cells per satellite. \\

        $K$
            & Maximum number of cells illuminated in one decision window. \\
        $t$
            & Decision-window index. \\
        $g_v$
            & Targeted geographic region. \\

        $\mu$
            & Target-region-to-satellite-and-cell mapping function. \\

        $\mathcal{B}$
            & Botnet of compromised user terminals. \\

        $\Gamma$
            & Attack traffic budget. \\

        $\mathbf{a}_t$
            & Cell-level attacker-generated traffic allocation for decision window $t$. \\

        $\pi_{s,t}$
            & Scheduling function used by satellite $s$ in decision window $t$. \\

        $\mathcal{F}_t$
            & Network-level source--destination flows realizing $\mathbf{a}_t$. \\

        $H$
            & Number of decision windows in the attack horizon. \\
    
        $\mathcal{T}$
            & Set of consecutive decision windows forming the attack horizon. \\

        $\mathbf{A}_{\mathcal{T}}$
            & Sequence of attacker-generated traffic allocations over $\mathcal{T}$. \\

        $\delta_v$
            & Victim-cell exclusion outcome. \\

        $\rho_v$
            & Attack success rate. \\

        \bottomrule
    \end{tabularx}
\end{table}

\subsection{System Model}
\label{sec:janus-system-model}

Let $\mathcal{S}$ denote the set of satellites.
Each satellite $s\in\mathcal{S}$ has $N$ BH cells, denoted by $\mathcal{C}_{s}=\{c_{s,1},\ldots,c_{s,N}\}$.
Let $K<N$ denote the maximum number of cells that may be illuminated during a BH decision window.
Let $g_v$ denote the targeted geographic region.
For each decision window $t$, the mapping function
\begin{equation}
\mu(g_v,t)=(s_t,c_{v,t})
\label{eq:victim-mapping}
\end{equation}
returns the satellite $s_t$ serving the targeted region and the BH cell $c_{v,t}$ to which it is mapped.
As LEO satellites continuously move relative to the Earth, the ground coverage of their cells changes over time.
Consequently, the same victim region may be served by different satellites or mapped to different cells across decision windows.
The attacker must therefore update this mapping before determining the attack traffic for each decision window.

The attacker controls a botnet $\mathcal{B}$ of compromised user terminals, each with a fixed geographic location and a bounded uplink rate.
To influence the demand of selected non-victim cells, the compromised terminals generate nominal user traffic toward ground regions served by those cells. 
This traffic alters the BH scheduler's view of demand by increasing the demand associated with those cells.

For decision window $t$, let $\mathbf{a}_t=[a_{t,1},\ldots,a_{t,N}]$ denote the additional attacker-generated traffic across the $N$ cells of the satellite serving the victim.
BH systems may employ different scheduling algorithms, such as rank-based rules or learning-based policies.
We denote by $\pi_{s,t}$ the scheduling function used by satellite $s$ in decision window $t$, which selects the cells to illuminate according to the implemented algorithm, the current network state, and the additional attacker-generated traffic vector $\mathbf{a}_t$, such that
\begin{equation}
\pi_{s,t}(\mathbf{a}_t) \subseteq \mathcal{C}_{s}.
\label{eq:illumination-selection}
\end{equation}
The number of cells selected by $\pi_{s,t}$ is bounded by $K$, such that $|\pi_{s,t}(\mathbf{a}_t)| \leq K$.
Accordingly, $\pi_{s,t}(\mathbf{0})$ represents the illumination selection under benign operation, while $\pi_{s,t}(\mathbf{a}_t)$ represents the selection after the attacker-generated traffic is incorporated.
For simplicity of notation, satellite and decision-window subscripts are omitted when they are clear from context.

\subsection{Attack Success Metrics}
\label{sec:janus-success-metrics}

Let $\mathcal{T}$ denote a set of consecutive BH decision windows forming the attack horizon.
Let $\mathbf{A}_{\mathcal{T}}=\{\mathbf{a}_t\}_{t\in\mathcal{T}}$ denote the sequence of attacker-generated traffic vectors over this horizon.
For each decision window $t\in\mathcal{T}$, the victim-specific attack outcome is defined as
\begin{equation}
\delta_v(t,\mathbf{a}_t)=
\begin{cases}
1, & \text{if } c_{v,t}\notin\pi_{s_t,t}(\mathbf{a}_t),\\
0, & \text{otherwise.}
\end{cases}
\label{eq:window-success}
\end{equation}
An attack with $|\mathcal{T}|=1$ is a single-window attack, while an attack with $|\mathcal{T}|>1$ is a multi-window attack spanning multiple consecutive decision windows.
A single-window attack is successful when $\delta_v(t,\mathbf{a}_t)=1$.

We denote the attack success rate over $\mathcal{T}$ by $\rho_v(\mathcal{T},\mathbf{A}_{\mathcal{T}})$, defined as
\begin{equation}
\rho_v(\mathcal{T},\mathbf{A}_{\mathcal{T}})
=
\frac{1}{|\mathcal{T}|}
\sum_{t\in\mathcal{T}}
\delta_v(t,\mathbf{a}_t).
\label{eq:attack-success-rate}
\end{equation}
The attack success rate measures the fraction of decision windows in which the BH cell serving the targeted geographic region is excluded from illumination.
Full-horizon success is achieved when $\rho_v(\mathcal{T},\mathbf{A}_{\mathcal{T}})=1$, indicating that the victim cell is excluded in every decision window of the attack horizon.

\subsection{Attack Planning and Execution}
\label{sec:janus-operation}

The planning method depends on the target scheduler.
For rank-based scheduling, JANUS computes a low-cost allocation that causes enough non-victim cells to outrank the victim, while for DRL it uses a budget-constrained evolutionary search guided by a surrogate scheduler.
Further planning details are provided in Appendix~\ref{app:attack-planning}.

A JANUS attack plan can use either iterative planning or horizon planning.
In iterative planning, each decision window is attacked independently when it is reached, without planning the complete attack in advance.
In horizon planning, the attacker jointly determines the attack traffic across the entire attack horizon while accounting for how the traffic and scheduling outcome in one decision window affect the conditions encountered in subsequent decision windows.
For both modes, network-level feasibility only ensures that the planned traffic can be realized by the botnet, not that the scheduler will produce the predicted allocation.

In iterative planning, the attacker treats each decision window as a separate planning problem.
For decision window $t$, the selected planning method uses the estimated scheduler inputs to determine the attacker-generated traffic $\mathbf{a}_t$ for the satellite serving the victim.
This traffic is then realized through the botnet as a set of source--destination flows $\mathcal{F}_t$, subject to compromised-terminal locations, per-terminal uplink limits, available routes, and residual link capacity.
The resulting flows are injected before the BH scheduler executes, and the same process is repeated independently for each subsequent decision window.
Algorithm~\ref{alg:janus-iterative} summarizes this procedure.

\begin{algorithm}[ht]
\caption{JANUS Iterative Planning}
\label{alg:janus-iterative}
\small
\begin{algorithmic}[1]
\Require Victim region $g_v$, botnet $\mathcal{B}$,
attack horizon $\mathcal{T}$, attack budget $\Gamma$,
and window-planning method $\mathsf{PlanWindow}$
\Ensure Attack outcome for each decision window

\For{each decision window $t\in\mathcal{T}$}
    \State Estimate the scheduler inputs for decision window $t$
    \State $(s_t,c_{v,t})\gets\mu(g_v,t)$
    \State $\mathbf{a}_t\gets
    \mathsf{PlanWindow}(t,s_t,c_{v,t},\Gamma)$
    \State $\mathcal{F}_t\gets
    \mathsf{Realize}(\mathbf{a}_t,\mathcal{B})$
    \If{$\mathcal{F}_t$ is infeasible}
        \State Record $\delta_v(t,\mathbf{0})$
        \State \textbf{continue}
    \EndIf
    \State Inject the traffic flows in $\mathcal{F}_t$
    \State Execute the BH scheduler
    \State Record $\delta_v(t,\mathbf{a}_t)$
\EndFor

\end{algorithmic}
\end{algorithm}

In horizon planning, the attacker treats the complete attack horizon
$\mathcal{T}=\{t_1,\ldots,t_H\}$ as a single planning problem.
The scheduler inputs and victim mapping are predicted for each decision window, and the attack traffic is jointly determined across the attack horizon while accounting for how the traffic and scheduling outcome in one decision window affect the conditions encountered in subsequent decision windows.
The resulting horizon-level attack plan is represented by $\mathbf{A}_{\mathcal{T}}$.
During execution, the planned traffic for each decision window is realized through the botnet as a set of source--destination flows $\mathcal{F}_t$, subject to compromised-terminal locations, per-terminal uplink limits, available routes, and residual link capacity.
Compromised terminals can be coordinated through control channels with specified transmission timing and rates, and synchronized using GNSS timing, as considered in prior LEO DDoS attacks~\cite{deng2025time,lu2024dosat}.
The resulting flows are injected in temporal order before the BH scheduler executes in each decision window.
Algorithm~\ref{alg:janus-horizon} summarizes this procedure.

\begin{algorithm}[ht]
\caption{JANUS Horizon Planning}
\label{alg:janus-horizon}
\small
\begin{algorithmic}[1]
\Require Victim region $g_v$, botnet $\mathcal{B}$,
attack horizon $\mathcal{T}$, attack budget $\Gamma$,
and horizon-planning method $\mathsf{PlanHorizon}$
\Ensure Attack outcome for each decision window

\State Predict the scheduler inputs over $\mathcal{T}$
\State Determine
$\{(s_t,c_{v,t})\}_{t\in\mathcal{T}}$
using $\mu(g_v,t)$
\State $\mathbf{A}_{\mathcal{T}}\gets
\mathsf{PlanHorizon}(g_v,\mathcal{T},\Gamma)$

\For{each decision window $t\in\mathcal{T}$}
    \State $\mathbf{a}_t\gets\mathbf{A}_{\mathcal{T}}[t]$
    \State $\mathcal{F}_t\gets
    \mathsf{Realize}(\mathbf{a}_t,\mathcal{B})$
    \If{$\mathcal{F}_t$ is infeasible}
        \State Record $\delta_v(t,\mathbf{0})$
        \State \textbf{continue}
    \EndIf
    \State Inject the traffic flows in $\mathcal{F}_t$
    \State Execute the BH scheduler
    \State Record $\delta_v(t,\mathbf{a}_t)$
\EndFor

\end{algorithmic}
\end{algorithm}

%% file: sections/Results.tex
\section{Results}
\label{sec:results}

Our evaluation aims to establish the feasibility of JANUS and characterize its potential to cause targeted service degradation.
We first examine whether adversarial traffic can manipulate individual BH scheduling decisions and whether this manipulation can be sustained over consecutive decision windows.
We then evaluate the resulting service degradation and the speed at which the service recovers after the attack ends.
Finally, we examine the resources required to carry out the attack and its robustness across different system configurations, DRL policies, and attacker-knowledge assumptions.

JANUS is implemented as an extension of ICARUS~\cite{giuliari2021icarus} with dynamic BH scheduling and adversarial traffic injection.
Unless stated otherwise, we use a Starlink G1-like constellation with GDP-weighted traffic under a nominal traffic load~\cite{jain2013b4,liu2024capa}.
Each satellite footprint contains $N=19$ candidate cells, of which at most $K=5$ are illuminated, with beam allocation decision updated every 20 ms, consistent with scheduling time scales used in prior BH studies~\cite{zhu2024beam,shuang2024joint}.
Unserved traffic remains queued for at most $L_{\mathrm{TTL}}=15$ decision windows (300~ms), providing a conservative upper bound relative to delay-sensitive LEO studies while remaining within standardized 5G packet-delay budgets~\cite{xu2025service,ahsan2025flexible,3gpp23501}, and each compromised terminal is limited to an upload rate of $25$~Mbps~\cite{starlinkuploadlimit}.
Victim cases are sampled across multiple decision windows to capture different traffic, coverage, and scheduling states.
Our single-window evaluation includes 1,000 victim cases sampled across 10 decision windows, while the multi-window evaluation uses 250 victim cases per attack horizon sampled across different starting windows.

Detailed constellation, traffic-generation, routing, victim-selection, and metric parameters are provided in Appendix~\ref{app:experimental-details}.

\subsection{Single-Window Attack Feasibility}
\label{sec:results-single}

We first evaluate whether JANUS can manipulate a single BH scheduling decision and prevent service to a target cell.
This experiment establishes the feasibility of demand manipulation before considering attacks sustained across multiple decision windows. 
A window is eligible when the victim is reachable, has traffic demand, and is mapped to a satellite cell.
We do not further restrict eligibility based on whether the victim would be selected without attack traffic, since the attacker constructs the attack without knowing this in advance.
As a result, some eligible victims are already unselected without attack traffic and contribute to the measured ASR without representing a change caused by JANUS.
We therefore use the fraction of victims unselected under benign operation as a baseline for interpreting the attack results.
Across the evaluated victim sets, this baseline is $25.7\%$ for KMAX and $25.11\%$ for DRL.
Additional details on victim selection are provided in Appendix~\ref{app:victim-sampling}.

\begin{figure}[ht]
    \centering
    \includegraphics[width=\columnwidth]{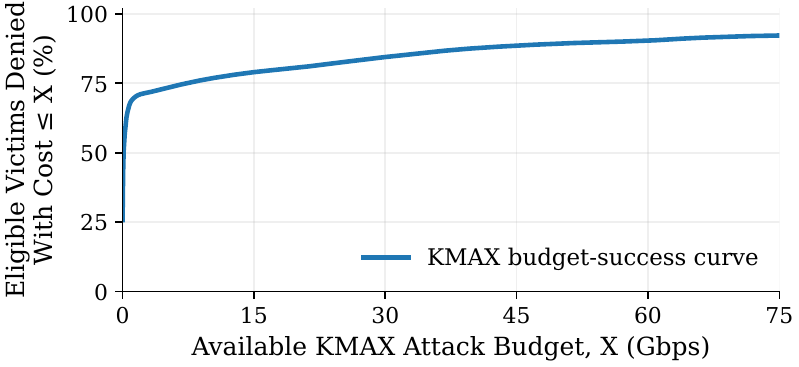}
    \caption{Resource-bounded feasibility of single-window JANUS against KMAX.
    The curve reports the percentage of eligible victims that can be excluded with an available attack budget of at most $X$.}
    \label{fig:single-window-kmax-cost}
\end{figure}

We begin with KMAX, whose explicit ranking rule allows us to directly determine the additional traffic required to displace the victim from the selected set.
JANUS successfully excludes the victim in $98.73\%$ of eligible cases.
Figure~\ref{fig:single-window-kmax-cost} shows the fraction of victims that can be excluded under different available traffic budgets.

The median traffic requirement for a successful KMAX attack is $0.44$~Gbps.
The 90th-percentile requirement reaches 65.90~Gbps, indicating that some target states require substantially greater resources.
Depending on network capacity, these requirements may exceed available ISL capacity or require traffic to be injected over preceding decision windows.
These results demonstrate both the feasibility of manipulating a single KMAX scheduling decision and the relatively low traffic required for a large fraction of targets.

We next evaluate JANUS against a learned DRL scheduler, where the relationship between cell demand and beam selection is less explicit.
JANUS achieves an ASR of $67.43\%$ with an attack budget of only $0.2$~Gbps, increasing to $92.09\%$ at $5$~Gbps.

Across both KMAX and DRL, the required traffic can be generated by botnets ranging from a few dozen up to just a few thousand compromised terminals.
Many of the evaluated attacks fall within the range of tens to hundreds of terminals, while more demanding target states can require larger botnets reaching into the low thousands.
These results show that JANUS can be carried out with relatively modest botnet resources for many targets, while more difficult cases require the attacker to scale to larger distributed botnets.

Overall, the single-window results show that JANUS can manipulate individual BH scheduling decisions against both rank-based and learned schedulers.
For many targets, this manipulation can be achieved with relatively small amounts of injected traffic, demonstrating the feasibility of targeted demand manipulation without requiring large traffic volumes.

\subsection{Continuous Denial and Service Degradation}
\label{sec:results-continuous}

Having established that JANUS can manipulate a single BH scheduling decision, we next ask whether this manipulation can be sustained across consecutive decisions and how sustained attacks affect victim service.
We evaluate multi-window attacks over horizons of up to $H=15$ decision windows, where $H=15$ matches the configured packet TTL.
Because each illumination decision changes the queue state observed in subsequent windows, multi-window attacks introduce temporal dependencies that are absent from the single-window setting.
We retain victims that are initially unselected under benign operation, since their scheduling priority can increase in subsequent decisions due to accumulated demand or fairness considerations.
We evaluate both iterative planning, which replans the attack before each decision, and horizon planning, which jointly determines the attack allocations across the complete horizon.
We measure service impact as throughput degradation relative to paired benign executions, $D_{\mathrm{thr}}=1-\sum_i S_i^{\mathrm{attack}}/\sum_i S_i^{\mathrm{benign}}$, over the attack-active windows.

We first evaluate KMAX to determine whether JANUS can sustain denial across consecutive decisions.
For $H=15$, JANUS achieves a mean ASR of $98.64\%$, while $94.8\%$ of victims are excluded in every decision of the attack horizon.
This near-continuous exclusion translates into near-complete service loss, reducing victim throughput by $99.7\%$ relative to the paired benign executions.
These results show that JANUS can extend individual scheduling manipulations into sustained denial across the complete attack horizon.

\begin{figure}[ht]
    \centering

    \includegraphics[width=\columnwidth]{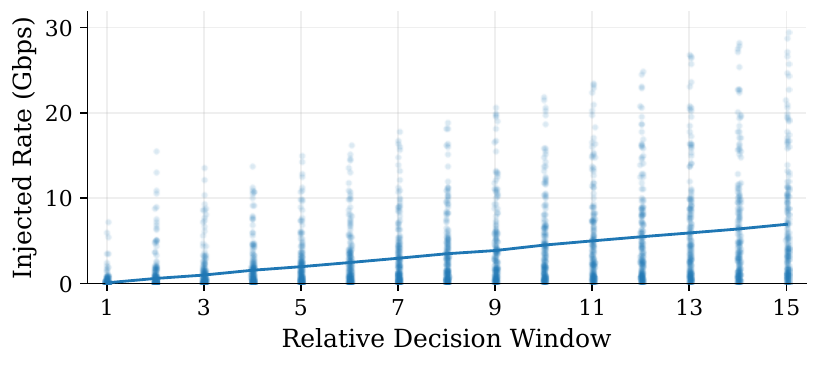}

    \includegraphics[width=\columnwidth]{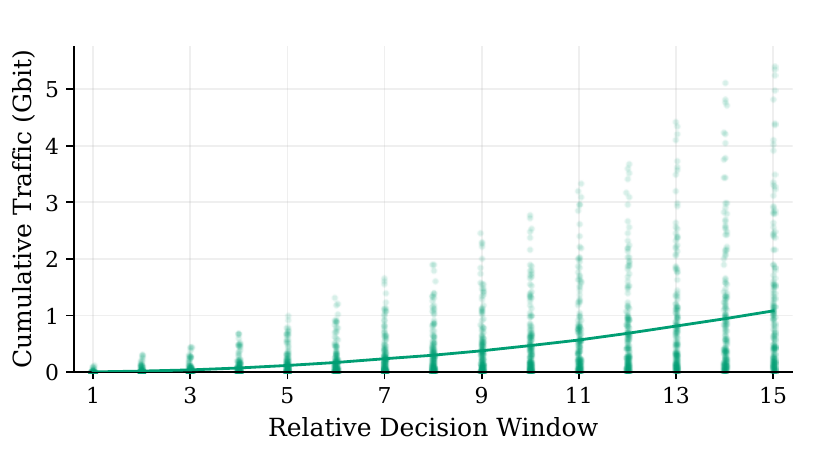}

    \caption{Resources required to sustain KMAX exclusion over $H=15$.
    The top panel shows the injected rate in each decision window, and the bottom panel shows cumulative injected traffic over the attack horizon.
    Dots show victim trials after per-window 75th-percentile filtering; the solid line shows the median.}
    \label{fig:kmax-horizon-cost}
\end{figure}

Sustaining this level of denial, however, becomes progressively more expensive.
Each missed service opportunity can increase the victim's scheduling priority as its queued traffic accumulates or ages, requiring the attacker to inject additional traffic into competing cells to keep the victim excluded.
Figure~\ref{fig:kmax-horizon-cost} shows the median injected rate increasing from less than $1$~Gbps to approximately $7.5$~Gbps over the attack horizon.
At the configured per-terminal uplink limit, this corresponds to the median KMAX attack growing from roughly a few dozen compromised terminals in the initial decisions to a few hundred by the end of the horizon.
This shows that sustaining near-continuous denial becomes progressively more expensive.

\begin{figure}[ht]
    \centering
    \includegraphics[width=\columnwidth]{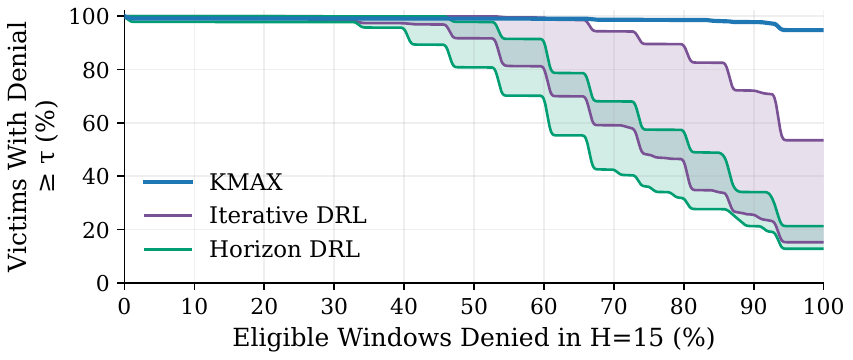}
    \caption{Victim-level denial persistence over the $H=15$ attack horizon.
    At threshold $\tau$, each curve reports the fraction of eligible victims whose per-victim ASR is at least $\tau$.
    The shaded DRL regions span the minimum and maximum empirical exceedance curves across the evaluated attack budgets.}
    \label{fig:denial-exceedance-combined}
\end{figure}

We next evaluate DRL, where the learned scheduling policy makes attack planning more challenging.
Despite this, JANUS remains effective across consecutive decisions.
Under iterative planning, JANUS achieves $74.23$--$92.66\%$ ASR across the evaluated budgets at $H=15$, with victim throughput degradation reaching $77.61$--$81.07\%$.
Similar results are observed across the evaluated attack horizons, with detailed per-budget results reported in Table~\ref{tab:app-recovery-by-horizon}.
Figure~\ref{fig:denial-exceedance-combined} shows the corresponding victim-level denial persistence.
These results show that substantial service degradation can be sustained against DRL across consecutive decisions.

We next evaluate whether JANUS can plan the attack jointly across the complete horizon.
At $H=15$, horizon planning remains effective, achieving $68.35$--$79.81\%$ ASR across the evaluated budgets.
Compared with iterative planning, the ASR is lower, but throughput degradation is higher, reaching approximately $83.41\%$.
This occurs because horizon planning accounts for the evolution of the victim state across decisions, allowing it to trade individual exclusions for greater cumulative service loss over the complete horizon.
These results demonstrate the feasibility of horizon-wide attack planning, and we leave further optimization of this method for future work.
Figure~\ref{fig:denial-exceedance-combined} shows the corresponding victim-level denial distribution.

Finally, we examine how long the service impact persists after the attack ends.
We continue each execution for 15 post-attack decisions and consider the victim recovered when its backlog is at most $105\%$ of the paired benign backlog, and its received service is at least $90\%$ of benign service.
Executions that do not satisfy both conditions within this period are treated as right-censored.

\begin{figure}[ht]
    \centering
    \includegraphics[width=\columnwidth]{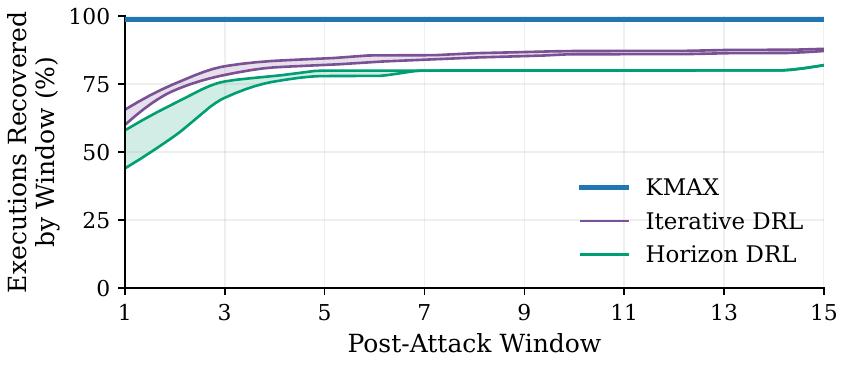}
    \caption{Post-attack recovery following $H=15$ JANUS attacks. Curves report the cumulative fraction of executions recovered by each post-attack decision. DRL bands span the evaluated budgets.}
    \label{fig:recovery-over-time}
\end{figure}

\input{tables/constellation_configurations}

Figure~\ref{fig:recovery-over-time} shows that although most DRL executions recover within the first few post-attack decisions, a non-negligible fraction remains degraded throughout the complete observation period.
Following iterative $H=15$ attacks, approximately $12.5\%$ of DRL executions do not recover within this period, increasing to $18\%$ following horizon-planned attacks.
In contrast, only $1.2\%$ of KMAX executions remain degraded, indicating substantially faster recovery under KMAX.

\subsection{Robustness Across System and Attacker Variations}
\label{sec:results-variations}

Having established JANUS under the baseline configuration, we next broaden the evaluation to better understand its feasibility across a wider range of system and scheduler conditions.
We vary the constellation profile, BH configuration, and decision-window duration, and evaluate independently trained DRL policies with different reward formulations.
Additional configuration details and experimental results are reported in Appendix~\ref{app:additional-results}.

\noindent\textbf{Effectiveness across constellation profiles.}
We first evaluate whether JANUS remains effective across different constellation configurations.
In addition to the baseline Starlink G1-like configuration~\cite{fcc_starlink1}, we consider a denser Starlink-like deployment~\cite{fcc_starlink2} and a OneWeb-like configuration~\cite{fcc_oneweb}.
These profiles are representative approximations chosen to capture differences in satellite density, altitude, and inclination rather than reproduce the corresponding commercial systems exactly.

Table~\ref{tab:constellation-variations} shows that the attack remains effective across all three constellation profiles.
Over $H=15$, DRL achieves $77.82\%$--$83.60\%$ mean ASR across the evaluated constellation profiles.
KMAX is less sensitive to the constellation variation, maintaining $97.00\%$--$99.12\%$ ASR across the single-window and $H=15$ experiments.

The resulting service impact is substantial across all configurations.
DRL throughput degradation ranges from $72.73\%$ to $86.72\%$ across the evaluated constellation and budget combinations, while KMAX reaches $99.68 \pm 0.08\%$ throughput degradation.

The variation in DRL effectiveness shows that the constellation configuration influences the conditions under which the attack is carried out.
Differences in satellite density and footprint characteristics change which cells compete with the victim and the traffic mapped to those cells, resulting in different scheduling conditions for the attacker.
Despite these differences, JANUS remains effective across all three evaluated profiles, showing that the attack is not specific to the baseline Starlink G1-like configuration.

\noindent\textbf{Sensitivity to BH configuration.}
We next examine how the BH configuration affects JANUS effectiveness and attack cost.
Using KMAX, we vary the number of candidate cells $N\in\{19,37,61\}$ and the number of simultaneously illuminated cells $K\in\{4,5,10\}$, spanning configurations considered in prior BH studies~\cite{deng2024satellites,hu2019deep,lyu2023beam,shuang2024joint}.
Table~\ref{tab:nk-sensitivity} reports the $H=15$ ASR and median average injected rate, normalized to the baseline $N=19,\ K=5$.

JANUS remains similarly effective across configurations: normalized ASR stays within approximately 1\% of the baseline except for $N=19,\ K=10$, where it remains 94.91\%.
In contrast, the median average injected rate ranges from 47.30\% to 213.20\% of the baseline, showing that $N$ and $K$ primarily affect attack cost rather than susceptibility to manipulation.
Full results for $H=5,10,15$, single-window ASR, and median and 90th-percentile costs are reported in Appendix Table~\ref{tab:app-nk-sensitivity-full}.

\begin{table}[ht]
    \centering
    \small
    \caption{Sensitivity of KMAX attacks to the BH configuration
    $N$ and $K$.
    Each cell reports normalized $H=15$ ASR / median average injected
    rate, relative to the baseline configuration $N=19,\ K=5$, which is
    normalized to $100\%/100\%$.}
    \label{tab:nk-sensitivity}
    \begin{tabular}{c|ccc}
        \toprule
        $N$ &
        $K=4$ &
        $K=5$ &
        $K=10$ \\
        \midrule

        19 &
        99.7\% / 89.8\% &
        \textbf{100\% / 100\%} &
        94.9\% / 213.2\% \\

        37 &
        100.9\% / 49.1\% &
        100.0\% / 93.2\% &
        100.2\% / 154.5\% \\

        61 &
        100.3\% / 47.3\% &
        100.8\% / 69.2\% &
        99.3\% / 95.5\% \\

        \bottomrule
    \end{tabular}
\end{table}

\noindent\textbf{Sensitivity to decision-window duration.}
We additionally evaluate JANUS over a 15-s period to examine whether the attack remains effective when substantially more traffic accumulates between scheduling decisions.
JANUS remains effective over the full period, with KMAX achieving 95.87\% ASR and DRL reaching up to 93.5\% ASR.
The 15-s horizon is consistent with the terminal-to-satellite assignment interval observed in Starlink~\cite{tanveer2023making}, showing that JANUS can sustain its effect over a practically relevant service period.
Full results are reported in Appendix~\ref{app:additional-results}.

\noindent\textbf{Generalization across DRL policies.}
Having shown that JANUS remains effective across different system configurations, we next examine whether its effectiveness depends on the particular learned DRL policy used in the baseline evaluation.
We train three additional DRL policies in the same simulation environment: one retains the primary reward formulation~\cite{xie2025multi,lin2022dynamic} with a different initialization, one introduces a drop penalty~\cite{liang2024multiagent}, and one uses fixed throughput normalization~\cite{chen2025distributed,zhao2025demand}.

\begin{table}[ht]
    \centering
    \small
    \caption{JANUS effectiveness across independently trained DRL policies.
    Values report the ASR range between attack budgets of $0.2$ and $5$~Gbps.}
    \label{tab:drl-policy-variations}
    \begin{tabular}{lcc}
        \toprule
        DRL policy &
        Single-window ASR &
        $H=15$ ASR \\
        \midrule

        Primary DRL
        & $67.43-92.09\%$
        & $72.62$--$91.88\%$ \\
        
        Same-Reward
        & $70.30-93.00\%$
        & $69.83-94.85\%$ \\

        Drop-Penalty
        & $65.02-88.53\%$
        & $65.78-93.79\%$ \\

        Fixed-Norm
        & $67.66-89.56\%$
        & $67.20-92.24\%$ \\

        \bottomrule
    \end{tabular}
\end{table}

\begin{figure}[ht]
    \centering
    \includegraphics[width=\columnwidth]{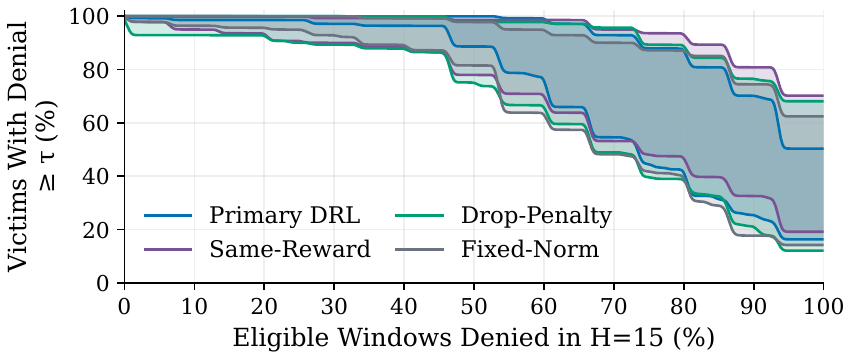}
    \caption{Denial persistence across independently trained DRL policies for $H=15$ attacks.
    At threshold $\tau$, each curve reports the fraction of eligible victims with per-victim ASR at least $\tau$.}
    \label{fig:drl-policy-exceedance}
\end{figure}

Table~\ref{tab:drl-policy-variations} shows that JANUS remains effective across all independently trained DRL policies.
At the lowest budget of $0.2$~Gbps, single-window ASR varies by only $5.28$ percentage points across policies, while $H=15$ ASR varies by $6.84$ points.
At $5$~Gbps, the spread narrows to $4.47$ points for single-window attacks and only $2.97$ points for $H=15$.
Thus, the independently trained policies exhibit similar vulnerability, with sustained attack effectiveness becoming particularly consistent at higher budgets.
Figure~\ref{fig:drl-policy-exceedance} shows that this similarity is also visible at the victim level, with closely aligned denial distributions across the four policies.

These results indicate that JANUS is not specific to a particular training initialization or reward formulation.

\subsection{Black-Box Attack Using Surrogates and External Observations}
Having established the feasibility of JANUS across the evaluated system and scheduler variations, we finally consider a black-box attack setting based on information that can be obtained externally.
The attacker derives system characteristics from publicly available documentation and estimates traffic patterns by observing network activity over time, while using independently trained DRL policies as surrogate models for the deployed scheduler.
We evaluate whether this externally derived information is sufficient to construct an effective $H=15$ attack against a fixed target DRL scheduler.

\begin{table}[ht]
    \centering
    \small
    \caption{JANUS $H=15$ ASR against a DRL scheduler using externally estimated traffic information and different models for attack construction.
    Values report mean ASR $\pm$ standard deviation across the four attack budgets of $0.2$, $0.5$, $2$, and $5$~Gbps.}
    \label{tab:drl-surrogate-variations}
    \begin{tabular}{lc}
        \toprule
        Attack construction & $H=15$ ASR \\
        \midrule
        Target DRL (reference) & $61.58 \pm 2.25\%$ \\
        Surrogate BASE & $59.43 \pm 2.18\%$ \\
        Surrogate DROP & $59.43 \pm 2.06\%$ \\
        Surrogate FIX\_NORM & $58.64 \pm 1.84\%$ \\
        \bottomrule
    \end{tabular}
\end{table}

Table~\ref{tab:drl-surrogate-variations} shows that JANUS remains effective when attacks are constructed using estimated traffic information and independently trained surrogate models.
Using the target DRL under the same traffic estimate provides a reference ASR of $61.58\%$, while the three surrogate models achieve a mean ASR of $59.17\%$, only $2.41$ percentage points lower.
The service impact also remains substantial, with throughput degradation ranging from $76.02\%$ to $77.07\%$ across the surrogate models and budgets, compared with approximately $81$--$84\%$ in the preceding evaluations.
These results show that a separately trained surrogate can approximate the deployed scheduler sufficiently well to retain similar attack effectiveness using estimated traffic information.
We also evaluate post-attack recovery for the black-box attacks, which remains similar to the preceding iterative evaluation.
At $H=15$, $13.48$--$13.91\%$ of executions remain right-censored, with detailed results reported in Appendix Table~\ref{tab:app-blackbox-recovery}.

\begin{figure}[ht]
\centering
\includegraphics[width=\columnwidth]{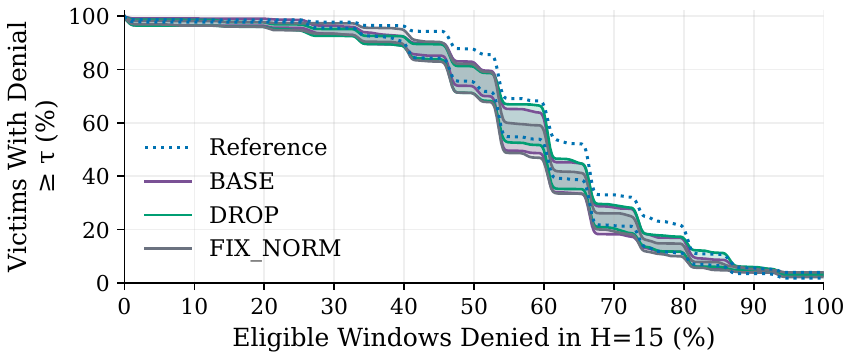}
\caption{Victim-level denial for $H=15$ attacks constructed using independently trained surrogate models and externally estimated traffic information.
The target scheduler remains fixed across all executions.
At threshold $\tau$, each curve reports the fraction of eligible victims with per-victim ASR at least $\tau$.}
\label{fig:drl-surrogate-transfer}
\end{figure}

Figure~\ref{fig:drl-surrogate-transfer} shows a similar result at the victim level, with closely aligned denial distributions across the three surrogate models.
These results show that JANUS remains effective in the black-box setting using surrogate models and estimated traffic information.
Overall, the variation experiments demonstrate the feasibility of the attack across different system configurations, schedulers, and attacker-knowledge settings.

%% file: tables/constellation_configurations.tex
\begin{table*}[!ht]
    \centering
    \small
    \caption{JANUS effectiveness across constellation configurations over the \(H=15\) attack horizon.
    DRL values report ASR at attack budgets from \(0.2\) to \(5\)~Gbps.}
    \label{tab:constellation-variations}

    \begin{tabular}{lccccc|cc}
    \toprule

    \multicolumn{6}{c|}{Constellation configuration}
    & \multicolumn{2}{c}{\(H=15\) ASR} \\

    Configuration
    & Planes
    & Sats/plane
    & Total sats
    & Altitude
    & Inclination
    & DRL
    & KMAX \\

    \midrule

    Starlink G1
    & 72
    & 22
    & 1,584
    & 550~km
    & \(53^\circ\)
    & \(72.62-91.88\%\)
    & 98.64\% \\

    Starlink Dense
    & 120
    & 30
    & 3,600
    & 530~km
    & \(53^\circ\)
    & \(66.38-86.67\%\)
    & 98.48\% \\

    OneWeb
    & 12
    & 49
    & 588
    & 1,200~km
    & \(87.9^\circ\)
    & \(75.51-89.80\%\)
    & 97.00\% \\

    \bottomrule
    \end{tabular}
\end{table*}

%% file: sections/Mitigation.tex
\section{Mitigation}
\label{sec:mitigation}

In this section, we examine how JANUS can be mitigated while retaining demand responsiveness as an input to beam-allocation decisions.
Using the same simulator and default configuration as in Section~\ref{sec:results}, we implement and evaluate several scheduler-side mitigation mechanisms that limit the attacker's influence on scheduling decisions.
We also discuss additional mitigation directions that could reduce the attacker's influence, increase the resources required to mount the attack, or facilitate detection and containment.

JANUS exploits the use of traffic demand in beam-allocation decisions.
Removing demand from the scheduling process would reduce this attack surface, but would also sacrifice the adaptability that makes demand-responsive BH effective.
Our goal is therefore to preserve the adaptability of demand-aware scheduling while making it harder for the adversary to manipulate beam allocation through adversarially shaped traffic.

\begin{table}[th]
\centering
\small
\caption{Attack reduction over the 15-decision horizon using a 5~Gbps DRL budget and min-cost KMAX planning.
ASR reduction is relative to the paired undefended attack and is reported in percentage points (pp).
}
\label{tab:mitigation}

\begin{tabular}{lcc}
    \toprule
    Mitigation &
    Defended ASR &
    Reduction \\
    \midrule

    \multicolumn{3}{l}{
        \textbf{DRL}\hspace{0.8em}\textit{Undefended ASR: 91.91\%}
    } \\

    Tiered reservation ($R=2$) &
    \textbf{49.63\%} &
    \textbf{42.29 pp} \\

    Tiered + consecutive ($R=1$) &
    64.17\% &
    27.75 pp \\

    Tiered reservation ($R=1$) &
    65.71\% &
    26.21 pp \\

    Random reservation ($R=1$) &
    75.50\% &
    16.42 pp \\

    Hard reservation ($R=1$) &
    77.45\% &
    14.46 pp \\

    EMA smoothing ($\alpha=0.5$) &
    79.41\% &
    12.50 pp \\

    Consecutive-service limit &
    88.74\% &
    3.18 pp \\

    \midrule

    \multicolumn{3}{l}{
        \textbf{KMAX}\hspace{0.8em}\textit{Undefended ASR: 99.75\%}
    } \\

    Tiered reservation ($R=2$) &
    \textbf{23.51\%} &
    \textbf{76.24 pp} \\

    Tiered reservation ($R=1$) &
    40.81\% &
    58.94 pp \\

    Random reservation ($R=1$) &
    43.82\% &
    55.93 pp \\

    Tiered + consecutive ($R=1$) &
    55.33\% &
    44.43 pp \\

    Consecutive-service limit &
    56.01\% &
    43.74 pp \\

    EMA smoothing ($\alpha=0.5$) &
    72.04\% &
    27.71 pp \\
        
    Hard reservation ($R=1$) &
    79.37\% &
    20.38 pp \\

    \bottomrule
\end{tabular}
\end{table}

\subsection{Mitigation Strategies}
\label{sec:mitigation-strategies}

We evaluate several scheduler-side mechanisms that either constrain the resulting beam allocation or reduce the immediate influence of traffic changes on scheduling decisions.

\noindent\textbf{Randomized Reservation.}
Motivated by probabilistic BH scheduling~\cite{feng2023performance}, we introduce controlled randomization.
We reserve $R$ of the $K$ beams for randomly selected nonempty cells that were not selected by the original scheduler.
The remaining $K-R$ beams follow the scheduler's original allocation.
The randomized reservation gives each eligible cell a chance of illumination in every decision, bounding the expected time a cell remains unserved.
The remaining beams are allocated dynamically according to demand, allowing the scheduler to adapt resource allocation to changing network needs.

\noindent\textbf{Starvation-Aware Reservation.}
Inspired by prior fairness-aware BH scheduling~\cite{deng2024satellites}, we prioritize backlogged cells according to how long they have remained unserved.
Two variants are evaluated.
In the hard variant, reserved beams are assigned among cells whose consecutive unserved duration exceeds a configured threshold.
In the tiered variant, cells exceeding this threshold are considered first and, if none are available, the threshold is progressively relaxed until eligible cells are found.
This directs reserved beams toward cells experiencing prolonged periods without service, reducing the possibility of sustained exclusion.

We evaluate randomized, hard, and tiered reservations with different reservation sizes $R$.

\noindent\textbf{Consecutive-Service Limit.}
Related round-robin BH schemes distribute illumination across cells to promote service fairness~\cite{deng2024satellites}.
We similarly limit how long a cell can remain continuously illuminated.
A cell that has been illuminated for $\tau_s$ consecutive decisions is temporarily excluded from selection.
We evaluate this mechanism independently with $\tau_s=3$ and together with tiered starvation-aware reservation using $\tau_u=3$ and $\tau_s=2$.
The combined mechanism limits both behaviors exploited during sustained attacks, prolonged exclusion of the victim and prolonged service of competing cells.

\noindent\textbf{Exponential Moving Average Smoothing.}
Building on prior use of exponential smoothing for satellite traffic management~\cite{bie2022queue}, we apply exponential moving average (EMA) smoothing to scheduler-visible demand.
Rather than using only the current demand state for beam selection, the scheduler combines the current value with its previously smoothed state.
Here, $\alpha$ controls the weight assigned to the current demand observation relative to the previously smoothed value.
We use $\alpha=0.5$ in our evaluation, giving equal weight to the current and historical demand state.

\subsection{Mitigation Effectiveness}
\label{sec:mitigation-effectiveness}

Table~\ref{tab:mitigation} shows that, among the evaluated mechanisms, those that directly modify beam allocation provide the largest reductions in JANUS effectiveness.
Tiered reservation with $R=2$ achieves the largest reduction in ASR among the evaluated mechanisms, reducing ASR by 42.29 percentage points for DRL and 76.24 percentage points for KMAX.
This reduces the portion of the allocation determined directly by demand-driven scheduling, while the tiered policy directs these beams toward cells that have remained unserved for longer periods, and directly counteracts the sustained exclusion exploited by JANUS.
The hard variant is less effective because it only applies when cells exceed the configured starvation threshold, whereas the tiered mechanism can progressively relax this threshold when necessary.

EMA smoothing provides more moderate reductions, decreasing ASR by 12.50 percentage points for DRL and 27.71 percentage points for KMAX.
Unlike the reservation and consecutive-service mechanisms, smoothing changes how quickly injected traffic affects scheduler-visible demand but does not directly constrain the resulting allocation.

The effectiveness of the remaining mechanisms varies between the two schedulers.
Randomized reservation reduces ASR by 16.42 percentage points for DRL and 55.93 percentage points for KMAX, while the consecutive-service limit reduces ASR by only 3.18 percentage points for DRL but 43.74 percentage points for KMAX.
The combined tiered and consecutive mechanism similarly produces different reductions across the two schedulers.
These differences show that mitigation effectiveness depends on how the scheduler maps demand state to beam-selection decisions, and defenses should therefore be evaluated for the specific scheduling policy being deployed.

Overall, among the evaluated mechanisms, defenses that directly constrain beam allocation provide larger reductions in JANUS effectiveness than smoothing the demand signal.

\subsection{Additional Mitigation Directions}
\label{sec:mitigation-additional}

Beyond the evaluated scheduler-side mechanisms, we suggest several additional directions for reducing an attacker's influence, increasing the resources required to manipulate beam allocation, or detecting manipulation before it persists.
We leave their experimental evaluation and further analysis of their implications to future work.

\noindent\textbf{Demand-Side Restrictions.} 
Prior work has considered finite buffering and admission control for managing traffic and resources in satellite networks~\cite{feng2023performance,dazhi2024joint}. 
Building on these ideas, we outline several demand-side restrictions that could limit how strongly large traffic demand influences beam selection. 
A physical queue cap bounds the backlog for each satellite--cell, while a per-window admission cap limits how much newly arriving demand is considered for beam selection. 
A scheduler-visible demand cap instead bounds the demand exposed to the scheduler without modifying the underlying physical queue. 
These mechanisms could reduce the influence of concentrated injected traffic, although restrictive caps may also suppress legitimate demand during periods of high load.

\noindent\textbf{Geographic Ingress Limits.}
Prior work has explored access-side DDoS defenses in satellite networks~\cite{guo2022distributed}.
One potential direction for JANUS is to impose ingress limits across geographic source areas.
Across the evaluated attacks, attack traffic remains concentrated within a small number of source grids.
Per-grid ingress limits could therefore increase the geographic cost of the attack by requiring traffic to be distributed across additional source areas.

\noindent\textbf{Detection and Containment.} 
One potential direction is to detect JANUS through coordinated demand changes and their scheduling effects, building on network-wide DDoS defenses and satellite-network intrusion detection~\cite{xing2021ripple,he2025synthetic}.
Detection could consider persistent demand increases across coordinated cells, their association with reduced service or repeated non-selection, and the geographic concentration of the responsible traffic.
JANUS relies on legitimate user traffic, making the injected traffic difficult to distinguish from benign demand based on individual flows alone.
A detector could therefore combine temporal, geographic, and cross-cell signals with scheduling outcomes to identify coordinated demand manipulation.

\noindent\textbf{Multi-Satellite Coordination.}
Another direction is to coordinate beam allocation across overlapping satellites~\cite{lin2022multi}.
Dense LEO constellations can provide overlapping coverage, allowing the same geographic area to be served by multiple satellites.
Coordinating beam allocation across these satellites could reduce dependence on the scheduling decision of any single satellite and provide alternative service when one scheduler is exposed to manipulated demand~\cite{wang2024resource}.
Such coordination may therefore make sustained exclusion more difficult for an attacker to coordinate and carry out.

\noindent\textbf{Dynamic Beamforming.}
This technology provides a complementary approach by dynamically adapting beam positions, shapes, or coverage areas rather than relying on a fixed BH cell structure~\cite{tang2021optimization,honnaiah2021demand}.
Such reconfiguration makes it more difficult for the attacker to predict how beam coverage and allocation will evolve in response to changing network conditions.

\noindent\textbf{Robust Scheduling Policies.}
Another potential direction is to incorporate resistance to demand manipulation directly into the scheduling policy.
Building on robust RL techniques~\cite{zhang2020robust,bukharin2023robust}, learning-based schedulers could be trained on adversarially shaped demand in addition to benign traffic.
The training objective could also encourage service fairness and penalize persistent exclusion under manipulated demand.

%% file: sections/Discussion.tex
\section{Discussion}\label{sec:discussion}

Having established the feasibility of JANUS and evaluated several mitigation strategies, we now discuss its practical implications, factors that may affect its real-world effectiveness, and broader directions for defending dynamic BH systems.

In real-world LEO networks, operational variability may make the relationship between injected traffic and scheduler-visible demand less predictable.
Attacker-generated traffic may traverse different paths and reach the relevant cell queues at different times, making it harder to align the injected demand with the intended BH decision window.
Existing traffic controls, such as authentication, rate-limiting mechanisms, and operator-level anomaly detection, as well as practical coordination across geographically distributed compromised terminals, may further constrain the attacker's ability to generate and synchronize the required traffic.
The scheduler may also rely on internal state, prioritization rules, and operational safeguards that are not observable to the attacker.
Together, these factors can make attack planning and execution more difficult and may change the quantitative effectiveness of JANUS in practice.
However, they do not remove the underlying security concern when traffic demand remains an input to dynamic resource-allocation decisions.

The mitigation results highlight a tradeoff in protecting dynamic BH against demand manipulation.
Operators must balance resistance to manipulation against other objectives such as responsiveness to legitimate demand, resource efficiency, fairness, and service requirements.
The appropriate balance may therefore differ across operational settings.
Mitigations should therefore be designed and evaluated according to the specific needs of the network.

Future BH system development should consider resistance to demand manipulation as part of the design and planning process.
Approaches such as multi-satellite coordination, dynamic beamforming, and robustness-aware scheduler training can change how demand influences resource allocation and therefore how susceptible the system is to manipulation.
Their impact on both scheduling performance and security should therefore be understood when these mechanisms are designed and deployed.

The evaluated mitigations reduce JANUS effectiveness, but they also change the conditions under which the attack must operate.
An attacker who accounts for the deployed defense may recover some of this effectiveness at the cost of additional resources or more complex planning.
Future work should therefore evaluate mitigations against adaptive attack strategies and quantify how much they increase the cost of successful demand manipulation.

%% file: sections/Conclusion.tex
\section{Conclusion}
\label{sec:conclusion}

As LEO satellite networks scale and become an important component of future 6G systems, ensuring the security of their dynamic resource-allocation mechanisms becomes critical.
In this work, we introduced JANUS, a targeted DoS attack that exposes a previously overlooked vulnerability in beam hopping arising from the scheduler's reliance on observed traffic demand.
By generating coordinated, legitimate traffic in selected non-victim cells, an adversary can manipulate beam-selection decisions and redirect beam resources away from the victim without compromising the satellite or directly flooding the victim.
We evaluated JANUS against different schedulers under sustained attacks, demonstrating substantial service disruption with relatively limited attacker resources.
Our results show that demand-responsive beam allocation creates a security-critical attack surface at the scheduling layer.
We further evaluated scheduler-side mitigations, showing that their effectiveness varies across schedulers.
JANUS demonstrates that dynamic BH schedulers introduce a new attack surface for DoS, underscoring the need to incorporate adversarial robustness into the design of future scheduling mechanisms.

%% file: sections/Ethical.tex
\section{Ethical Considerations}
\label{sec:ethics}

JANUS is presented to expose a broader security risk in which an adversary can manipulate a resource-allocation scheduler through the inputs used to guide its decisions.
The purpose of this work is to raise awareness of this attack surface, study the conditions that make such manipulation possible, and provide an initial evaluation of potential mitigation strategies.
All experiments were conducted in simulation and did not interact with operational satellite infrastructure or affect real users.
By presenting this risk, we aim to encourage further research on secure scheduling, detection, and defense mechanisms for emerging LEO satellite networks.

%% file: sections/Appendix.tex
\appendix

\subsection{Attack-Planning Details}
\label{app:attack-planning}

This section provides additional implementation details for the attack-planning procedures used in our evaluation.
It supplements the main attack description with the configuration and search details used in the experiments.

\noindent\textbf{Rank-based planning.}
For KMAX, the scheduler ranks candidate cells according to their scheduler-visible backlog.
For each eligible non-victim cell, JANUS determines the additional traffic required for that cell to outrank the victim under the scheduler's deterministic ordering.
It then searches for a low-cost set of competing cells sufficient to displace the victim from the top-$K$ selected cells.
When two cells have equal backlog, the scheduler's deterministic cell ordering is used to resolve the tie, and the required traffic increment is adjusted accordingly.
The planner therefore searches for allocations that reduce the total cell-level traffic required for exclusion, subject to the scheduler constraints.
Network-level feasibility is evaluated separately when the resulting allocation is realized through the botnet.

\noindent\textbf{DRL planning.}
For the DRL scheduler, JANUS performs a budget-constrained evolutionary search over attacker-generated traffic allocations to non-victim cells.
Each candidate represents additional traffic injected in the current decision window and is constrained by the attack budget and the maximum number of attacked cells.
Candidates are evaluated using the attack-construction scheduler.
The evaluated configuration uses a population of 32 candidates for at most 80 generations and terminates once a candidate that excludes the victim is found.
The initial population combines single-cell full-budget candidates with randomized sparse allocations.
Subsequent generations retain the highest-ranked candidates and generate new candidates through crossover, mutation, and random restart.
For horizon planning, the attack allocations are computed jointly across the complete attack horizon, allowing the selected traffic allocation to vary across decision windows.
To account for future arrivals, we also experimented with different traffic predictors, including a history-based predictor derived from prior observations and an ensemble predictor that uses Monte Carlo simulation to sample possible future traffic according to the GDP-weighted traffic model.

\subsection{Experimental Details}
\label{app:experimental-details}

This appendix provides the experimental configuration, victim-sampling procedure, and metric definitions supporting the evaluation in Section~\ref{sec:results}.
Unless otherwise stated, experiments use the default configuration summarized in Table~\ref{tab:app-default-config}.

\begin{table}[ht]
    \centering
    \small
    \caption{Default JANUS experimental configuration.}
    \label{tab:app-default-config}
    \begin{tabular}{ll}
        \toprule
        Parameter & Value \\
        \midrule
        Constellation & Starlink G1-like \\
        Orbital planes & 72 \\
        Satellites per plane & 22 \\
        Total satellites & 1,584 \\
        Altitude & 550~km \\
        Inclination & $53^\circ$ \\
        Ground model & GDP-weighted geodesic grid \\
        Routing & Shortest path \\
        Candidate cells, $N$ & 19 \\
        Illuminated cells, $K$ & 5 \\
        Decision-window duration & 20~ms \\
        Packet TTL, $L_{\mathrm{TTL}}$ & 15 windows (300~ms) \\
        Beam-interference model & SINR-aware \\
        Per-terminal uplink limit & 25~Mbps \\
        DRL attack budgets & 0.2, 0.5, 2, 5~Gbps \\
        Single-window trials & 1,000 \\
        Multi-window trials & 250 per horizon \\
        Attack horizons, $H$ & 5, 10, 15 decision windows \\
        \bottomrule
    \end{tabular}
\end{table}

\noindent\textbf{Victim Sampling.}
\label{app:victim-sampling}

Single-window victims are sampled across $10$ decision windows and stratified according to the backlog distribution.
JANUS does not assume that the attacker knows whether a victim would be selected in the corresponding benign execution.
Because only $K=5$ of $N=19$ cells can be illuminated in each decision window, some sampled victims are naturally unselected even without attack traffic.

For the primary single-window evaluation, the benign non-selection rate is $25.11\%$ for the DRL victim cohort and $25.7\%$ for the KMAX cohort.
On the primary DRL cohort, JANUS achieves $67.43\%$ ASR at $0.2$~Gbps and $92.09\%$ at $5$~Gbps.
In our single-window paired executions, all victims that are unselected under benign operation remain unselected under attack; hence, the benign non-selection set is a subset of the attacked non-selection set.
To isolate the denial induced by JANUS beyond benign non-selection, we subtract the benign non-selection rate from the observed ASR and normalize by the remaining fraction:
\begin{equation}
\mathrm{ASR}_{\mathrm{induced}}
=
\frac{\mathrm{ASR}-b}{1-b},
\end{equation}
where $b$ denotes the benign non-selection rate.
This correction is used only for the single-window analysis and is not applied to continuous multi-window attacks, where benign and attacked scheduling trajectories evolve differently over time.
For the primary DRL cohort, where $b=25.11\%$, the observed ASRs of $67.43\%$ and $92.09\%$ at attack budgets of $0.2$ and $5$~Gbps correspond to attack-induced denial rates of $56.51\%$ and $89.44\%$, respectively.
For KMAX, the observed ASR of $98.73\%$ with $b=25.7\%$ corresponds to an attack-induced denial rate of $98.29\%$.
We additionally evaluate JANUS on independent $1{,}000$-case DRL cohorts with different benign non-selection rates, including the same $1{,}000$ victims used in the KMAX evaluation.

\input{tables/Recovery_table}
\noindent\textbf{DRL Training.}
\label{app:drl-training}

The DRL scheduler used in our simulation is trained with PPO under an SINR-aware service model.
Its scalar reward at decision window $t$, denoted $r_t$, balances normalized served traffic against a latency term derived from queue age:
\[ r_t=\beta T_t-(1-\beta)L_{\mathrm{age}}, \]
where $T_t$ is the normalized amount of traffic served during decision window $t$, and $L_{\mathrm{age}}$ is the normalized average age of queued traffic, used as a proxy for queueing latency because traffic that remains unserved accumulates age across consecutive decision windows.
The parameter $\beta$ controls the tradeoff between throughput and latency.
Training was configured for $50$ million environment steps.
The deployed scheduler uses the best validation checkpoint, selected after approximately $39.26$ million training steps.

\begin{table}[ht]
    \centering
    \small
    \caption{Training configuration of the DRL scheduler.}
    \label{tab:app-drl-training}
    \begin{tabular}{ll}
        \toprule
        Parameter & Value \\
        \midrule
        Algorithm & PPO \\
        Throughput weight, $\beta$ & 0.05 \\
        Latency weight, $1-\beta$ & 0.95 \\
        Learning rate & 0.0005 \\
        Rollout steps & 512 \\
        Parallel environments & 8 \\
        Samples per rollout & 4,096 \\
        Batch size & 512 \\
        Optimization epochs/update & 10 \\
        Discount, $\gamma$ & 0.99 \\
        GAE, $\lambda$ & 0.95 \\
        PPO clip range & 0.2 \\
        Value-function coefficient & 0.5 \\
        Entropy coefficient & 0 \\
        Maximum gradient norm & 0.5 \\
        Target KL & 0.03 \\
        Advantage normalization & Yes \\
        Optimizer & Adam \\
        \bottomrule
    \end{tabular}
\end{table}

Training uses Python~3.11.11, Stable-Baselines3~2.7.0, PyTorch~2.9.1+cu128, Gymnasium~1.2.2, and NumPy~1.26.4.
GPU acceleration is disabled.

\subsection{Recovery Across Attack Horizons}
\label{app:recovery-by-horizon}

Section~\ref{sec:results-continuous} focuses on recovery following $H=15$ attacks.
Here, we provide the corresponding results for shorter attack horizons and the complete horizon-separated recovery values.

\begin{figure*}[th]
    \centering

    \begin{minipage}{\columnwidth}
        \centering
        \includegraphics[width=\linewidth]{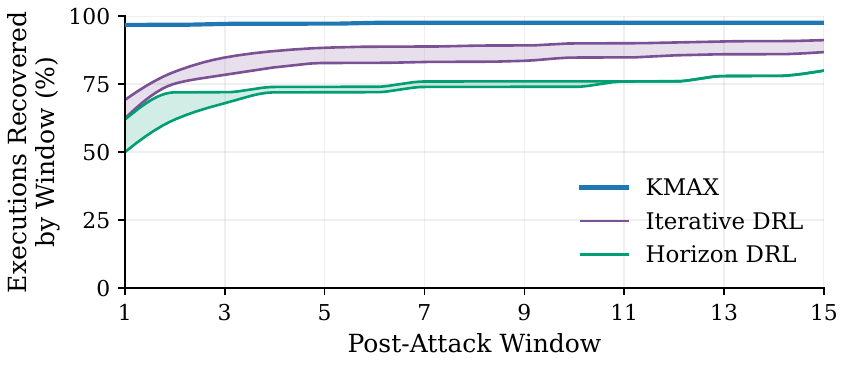}
        \footnotesize (a) $H=5$
    \end{minipage}
    \hfill
    \begin{minipage}{\columnwidth}
        \centering
        \includegraphics[width=\linewidth]{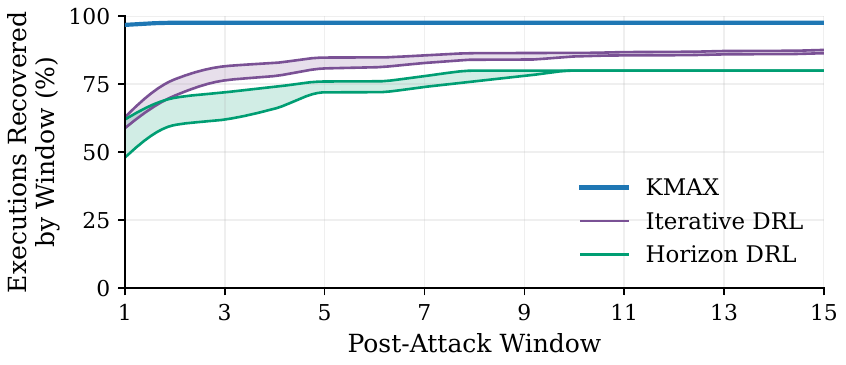}
        \footnotesize (b) $H=10$
    \end{minipage}

    \caption{Post-attack recovery following shorter JANUS attack horizons.
    Each curve reports the cumulative percentage of executions recovered by the corresponding post-attack decision for (a) $H=5$ and (b) $H=10$.
    DRL bands span the four evaluated attack budgets.}
    \label{fig:app-recovery-short-horizons}
\end{figure*}

Figure~\ref{fig:app-recovery-short-horizons} shows the post-attack recovery trajectories following $H=5$ and $H=10$ attacks.
Although the trajectories differ during the first few post-attack decision windows, they converge to similar levels by the end of the 15-window observation period.
For iterative DRL, $86.4\%$--$91.2\%$ of executions recover by the end of the observation period across the evaluated horizons and budgets.
Horizon planning reaches $80\%$ recovery following both $H=5$ and $H=10$, compared with $82\%$ following $H=15$.
KMAX recovers most rapidly, with at least $97.6\%$ of executions recovered across all evaluated horizons.

Table~\ref{tab:app-blackbox-recovery} shows that post-attack recovery remains similar across the evaluated surrogate models and attack budgets.
For $H=15$, approximately $86\%$ of executions recover within the $15$-window observation period, leaving only $13.48$--$13.91\%$ right-censored.
The close recovery ranges across surrogates are consistent with the main black-box results, indicating that surrogate mismatch has a limited effect on the persistence of the resulting service degradation.
\input{tables/Recovery_blackbox}

\subsection{Additional Results and Evaluations}
\label{app:additional-results}

\noindent\textbf{Sensitivity to $N$ and $K$.}

\begin{table*}[ht]
    \centering
    \small
    \caption{Normalized KMAX sensitivity to the beam-hopping configuration $N$ and $K$.
    Each value is reported relative to the corresponding baseline $N=19,\ K=5$, normalized to $100\%$.
    Cost values report the median/90th-percentile realized average injected rate.}
    \label{tab:app-nk-sensitivity-full}
    \resizebox{\textwidth}{!}{%
    \begin{tabular}{cc|ccc|ccc|ccc|ccc}
        \toprule
        \multirow{2}{*}{$N$} &
        \multirow{2}{*}{$K$} &
        \multicolumn{3}{c|}{Single window} &
        \multicolumn{3}{c|}{$H=5$} &
        \multicolumn{3}{c|}{$H=10$} &
        \multicolumn{3}{c}{$H=15$} \\
        \cmidrule(lr){3-5}
        \cmidrule(lr){6-8}
        \cmidrule(lr){9-11}
        \cmidrule(lr){12-14}
        &
        &
        ASR &
        Med. &
        P90 &
        ASR &
        Med. &
        P90 &
        ASR &
        Med. &
        P90 &
        ASR &
        Med. &
        P90 \\
        \midrule

        19 & 4 &
        100.47\% & 90.94\% & 71.05\% &
        99.84\% & 89.88\% & 64.81\% &
        99.71\% & 84.96\% & 60.00\% &
        99.68\% & 89.84\% & 62.44\% \\

        19 & 5 &
        \textbf{100\%} & \textbf{100\%} & \textbf{100\%} &
        \textbf{100\%} & \textbf{100\%} & \textbf{100\%} &
        \textbf{100\%} & \textbf{100\%} & \textbf{100\%} &
        \textbf{100\%} & \textbf{100\%} & \textbf{100\%} \\

        19 & 10 &
        94.26\% & 327.55\% & 197.64\% &
        95.26\% & 240.65\% & 178.12\% &
        94.76\% & 229.48\% & 180.45\% &
        94.91\% & 213.20\% & 168.15\% \\

        \midrule

        37 & 4 &
        100.96\% & 54.72\% & 53.26\% &
        101.25\% & 54.29\% & 33.47\% &
        101.04\% & 52.51\% & 42.65\% &
        100.95\% & 49.05\% & 38.28\% \\

        37 & 5 &
        100.95\% & 67.17\% & 69.35\% &
        100.80\% & 99.27\% & 44.37\% &
        100.51\% & 97.81\% & 61.72\% &
        100.04\% & 93.21\% & 59.01\% \\

        37 & 10 &
        100.10\% & 130.57\% & 98.63\% &
        100.38\% & 165.63\% & 100.41\% &
        100.22\% & 163.63\% & 99.79\% &
        100.21\% & 154.48\% & 96.71\% \\

        \midrule

        61 & 4 &
        101.08\% & 50.57\% & 43.80\% &
        100.83\% & 36.33\% & 28.19\% &
        100.63\% & 43.19\% & 29.27\% &
        100.35\% & 47.30\% & 30.09\% \\

        61 & 5 &
        101.08\% & 52.08\% & 51.03\% &
        101.25\% & 59.84\% & 29.20\% &
        101.00\% & 73.07\% & 39.07\% &
        100.85\% & 69.24\% & 37.33\% \\

        61 & 10 &
        100.16\% & 99.62\% & 88.33\% &
        99.62\% & 91.51\% & 62.92\% &
        99.45\% & 89.98\% & 74.72\% &
        99.35\% & 95.53\% & 76.57\% \\

        \bottomrule
    \end{tabular}%
    }
\end{table*}

Table~\ref{tab:app-nk-sensitivity-full} provides the complete KMAX sensitivity results across the evaluated beam-hopping configurations.
Attack effectiveness remains consistently high across the configurations: single-window ASR ranges from 93.06\% to 99.80\%, while $H=15$ ASR ranges from 93.86\% to 99.83\%.
The resource requirement is considerably more sensitive to the configuration.
Increasing $K$ generally raises the attack cost because more competing cells must be promoted to displace the victim, with the strongest effect observed for $N=19,\ K=10$.
Overall, the results indicate that changing the beam-hopping configuration has a substantially larger effect on the resources required to sustain JANUS than on its ability to manipulate KMAX.

\noindent\textbf{Antenna and Spatial Constraints}
\label{app:antenna-constraints}

\begin{figure*}[ht]
    \centering

    \begin{minipage}[t]{0.48\textwidth}
        \centering
        \includegraphics[width=\linewidth]{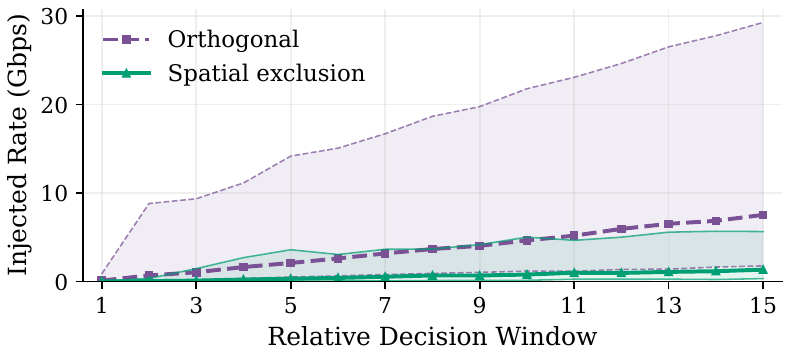}
        \footnotesize (a) Per-window injected rate
    \end{minipage}
    \hfill
    \begin{minipage}[t]{0.48\textwidth}
        \centering
        \includegraphics[width=\linewidth]{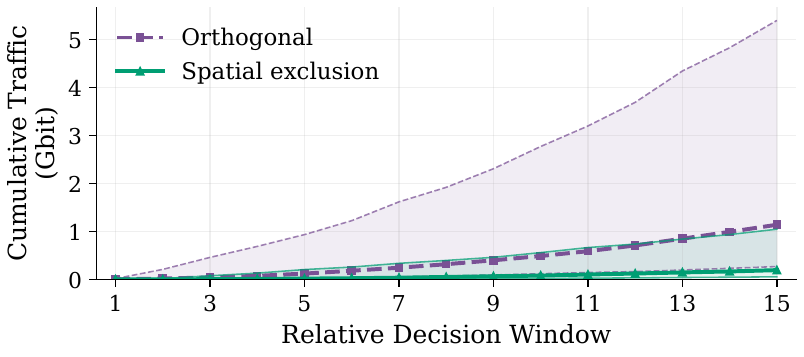}
        \footnotesize (b) Cumulative injected traffic
    \end{minipage}

    \caption{KMAX attack cost under different beam-feasibility models over the $H=15$ attack horizon.
    (a) Injected rate in each decision window.
    (b) Cumulative injected traffic.
    The baseline SINR-aware configuration is omitted because its resource distribution
    nearly overlaps the orthogonal configuration.}
    \label{fig:antenna-config-cost}
\end{figure*}

We additionally evaluate whether JANUS depends on the interference-management abstraction used by the BH scheduler.
All configurations use KMAX with the same minimum-cost attack planner, while varying how concurrent beam illumination is constrained.
\emph{Orthogonal KMAX} assumes that potentially interfering beams use orthogonal channel resources, omitting co-channel interference from the service model~\cite{honnaiah2023demand}.
\emph{SINR-aware KMAX}, used in our default configuration, permits concurrent illumination while accounting for interference through the resulting service rate~\cite{meng2025joint}.
Finally, \emph{Spatial-exclusion KMAX} imposes a hard feasibility constraint that prevents neighboring cells from being illuminated simultaneously, following the general principle of distance-constrained beam hopping~\cite{zhang2023system}.

Across all three configurations, JANUS remains highly effective.
Single-window ASR is $98.73\%$ under both Orthogonal and SINR-aware KMAX and $99.77\%$ under spatial exclusion.
Over the $H=15$ attack horizon, Orthogonal and SINR-aware KMAX achieve $98.64\%$ ASR with $94.78\%$ full-horizon exclusion, while spatial exclusion increases these values to $99.42\%$ and $97.83\%$, respectively.
Thus, changing the interference model does not eliminate the vulnerability in the evaluated setting.

The main difference appears in attack cost.
Figure~\ref{fig:antenna-config-cost} shows that Orthogonal and SINR-aware KMAX produce nearly identical cost traces, while spatial exclusion requires substantially less injected traffic throughout the attack horizon.
Under Orthogonal and SINR-aware KMAX, the median injected rate rises from below $1$~Gbps in the first attack decision to approximately $7.5$~Gbps in the final decision, and the median cumulative attack traffic reaches approximately $1.1$~Gbit.
Under spatial exclusion, the final median injected rate remains near $1.4$~Gbps, and the median cumulative traffic remains below $0.25$~Gbit.

This reduction follows from the structure of the spatial-exclusion rule.
When neighboring cells cannot be illuminated simultaneously, JANUS can promote cells whose selection is incompatible with serving the victim.
A relatively small amount of injected demand can therefore remove the victim from the feasible set.
In this setting, the hard spatial constraint makes KMAX less expensive to manipulate, even though all three antenna models remain highly vulnerable.

\noindent\textbf{Sensitivity to decision-window duration.}
To further evaluate the effect of longer decision windows, we increase the decision-window duration from 20~ms to 1~s while preserving the same traffic-to-capacity ratio within each window.
The resulting $H=15$ horizon spans 15~s, consistent with the terminal-to-satellite assignment interval observed in Starlink~\cite{tanveer2023making}.

\begin{table}[ht]
    \centering
    \small
    \caption{JANUS effectiveness with 1-s decision windows.
    Values report ASR over eligible decision windows.}
    \label{tab:window-duration}
    \begin{tabular}{llcc}
        \toprule
        \textbf{Scheduler} &
        \textbf{Budget} &
        \textbf{Single window} &
        \textbf{$H=15$} \\
        \midrule

        \multirow{4}{*}{DRL}
        & 0.2~Gbps & 82.2\% & 78.7\% \\
        & 0.5~Gbps & 88.2\% & 85.4\% \\
        & 2~Gbps   & 94.8\% & 91.4\% \\
        & 5~Gbps   & 96.5\% & 93.5\% \\

        \midrule

        KMAX
        & Minimize cost
        & 94.93\%
        & 95.87\% \\

        \bottomrule
    \end{tabular}
\end{table}

Table~\ref{tab:window-duration} shows that JANUS remains effective with 1-s decision windows.
Against the DRL scheduler, single-window ASR ranges from 82.2\% to 96.5\%, while over $H=15$ it ranges from 78.7\% to 93.5\%.
KMAX similarly maintains 94.93\% single-window ASR and 95.87\% ASR over $H=15$.
Compared with the 20-ms baseline, the longer decision window therefore has only a limited effect on attack effectiveness.

The longer the window, the more resources are required to manipulate each decision.
For single-window KMAX attacks, the median successful injected rate decreases from 0.44~Gbps at 20~ms to 0.20~Gbps at 1~s, corresponding to approximately 18 and 8 compromised terminals, respectively.
However, traffic accumulates for $50\times$ longer before each decision: over $H=15$, median cumulative KMAX traffic increases from approximately 1.1~Gbit to 20~Gbit.

%% file: tables/Recovery_table.tex
\begin{table*}[ht]
    \centering
    \small
    \caption{Post-attack recovery across attack horizons and planning
    strategies.
    Values report the cumulative percentage of executions recovered by
    post-attack decision windows 1, 3, and 15.
    DRL ranges span the four evaluated attack budgets.
    Right-censored is the fraction not recovered by window 15.}
    \label{tab:app-recovery-by-horizon}

    \begin{tabular}{lccccc}
        \toprule
        \textbf{Planner} &
        \textbf{Horizon} &
        \textbf{Window 1} &
        \textbf{Window 3} &
        \textbf{Window 15} &
        \textbf{Right-censored} \\
        \midrule

        \multirow{3}{*}{Iterative}
        & \(H=5\)
        & 62.4--69.2\%
        & 78.4--84.8\%
        & 86.8--91.2\%
        & 8.8--13.2\% \\

        & \(H=10\)
        & 58.8--62.8\%
        & 76.4--81.6\%
        & 86.4--87.6\%
        & 12.4--13.6\% \\

        & \(H=15\)
        & 60.0--65.6\%
        & 78.4--81.6\%
        & 87.2--88.0\%
        & 12.0--12.8\% \\

        \midrule

        \multirow{3}{*}{Horizon}
        & \(H=5\)
        & 50.0--62.0\%
        & 68.0--72.0\%
        & 80.0\%
        & 20.0\% \\

        & \(H=10\)
        & 48.0--62.0\%
        & 62.0--72.0\%
        & 80.0\%
        & 20.0\% \\

        & \(H=15\)
        & 44.0--58.0\%
        & 70.0--76.0\%
        & 82.0\%
        & 18.0\% \\

        \midrule

        \multirow{3}{*}{KMAX}
        & \(H=5\)
        & 96.8\%
        & 97.2\%
        & 97.6\%
        & 2.4\% \\

        & \(H=10\)
        & 96.8\%
        & 97.6\%
        & 97.6\%
        & 2.4\% \\

        & \(H=15\)
        & 98.8\%
        & 98.8\%
        & 98.8\%
        & 1.2\% \\

        \bottomrule
    \end{tabular}
\end{table*}

%% file: tables/Recovery_blackbox.tex
\begin{table*}[ht]
    \centering
    \small
    \caption{Post-attack recovery for black-box JANUS attack.
    Values report the cumulative percentage of executions recovered by post-attack decision windows 1, 3, and 15.
    Ranges span the evaluated attack budgets.
    Right-censored is the fraction not recovered by window 15.}
    \label{tab:app-blackbox-recovery}

    \begin{tabular}{lccccc}
        \toprule
        \textbf{Attack surrogate} &
        \textbf{Horizon} &
        \textbf{Window 1} &
        \textbf{Window 3} &
        \textbf{Window 15} &
        \textbf{Right-censored} \\
        \midrule

        \multirow{3}{*}{Drop-Penalty}
        & $H=5$
        & 58.77--61.84\%
        & 76.32--77.63\%
        & 85.09--85.53\%
        & 14.47--14.91\% \\

        & $H=10$
        & 58.08--61.14\%
        & 75.55--76.86\%
        & 85.15--85.59\%
        & 14.41--14.85\% \\

        & $H=15$
        & 59.13--61.74\%
        & 77.39--78.70\%
        & 86.09\%
        & 13.91\% \\

        \midrule

        \multirow{3}{*}{FIX NORM}
        & $H=5$
        & 57.89--60.09\%
        & 76.32--78.07\%
        & 85.09\%
        & 14.91\% \\

        & $H=10$
        & 56.33--60.70\%
        & 75.55--77.73\%
        & 85.15--85.59\%
        & 14.41--14.85\% \\

        & $H=15$
        & 54.35--62.17\%
        & 76.52--79.13\%
        & 86.09\%
        & 13.91\% \\

        \midrule

        \multirow{3}{*}{Independent Training}
        & $H=5$
        & 60.09--64.47\%
        & 76.75--79.82\%
        & 85.09--85.53\%
        & 14.47--14.91\% \\

        & $H=10$
        & 56.77--59.83\%
        & 75.55--77.29\%
        & 85.59--86.03\%
        & 13.97--14.41\% \\

        & $H=15$
        & 55.65--61.30\%
        & 76.96--78.70\%
        & 86.09--86.52\%
        & 13.48--13.91\% \\

        \bottomrule
    \end{tabular}
\end{table*}